\documentclass{elsart}

\usepackage[english]{babel}
\usepackage{graphicx,rotating,amssymb}
\def\ifmath#1{\relax\ifmmode #1\else $#1$\fi}%

\def\TeV{\ifmmode {\,\mathrm{ Te\kern -0.1em V}}\else
                   \textrm{Te\kern -0.1em V}\fi}%
\def\GeV{\ifmmode {\,\mathrm{ Ge\kern -0.1em V}}\else
                   \textrm{Ge\kern -0.1em V}\fi}%
\def\MeV{\ifmmode {\,\mathrm{ Me\kern -0.1em V}}\else
                   \textrm{Me\kern -0.1em V}\fi}%
\def\keV{\ifmmode {\,\mathrm{ ke\kern -0.1em V}}\else
                   \textrm{ke\kern -0.1em V}\fi}%
\def\eV{\ifmmode  {\,\mathrm{ e\kern -0.1em V}}\else
                   \textrm{e\kern -0.1em V}\fi}%

\newcommand{\pb}{\,{\rm pb}}
\newcommand{\fb}{\,{\rm fb}}

\newcommand{\fbi}{\, \fb^{-1}}

\newcommand{\cm}{\mathrm{cm}}

\newcommand{\micron}{\mu\mathrm{m}}

\newcommand{\mrad}{\mathrm{mrad}}
\newcommand{\lunit}{\mathrm{cm}^{-2}\mathrm{s}^{-1}}

\newcommand{\WW}    {\mathrm{W}^+\mathrm{W}^-}
\newcommand{\ee}    {\mathrm{e}^+\mathrm{e}^-}
\newcommand{\mumu}  {\mu^+\mu^-}

\newcommand{\bb}    {{\mathrm b\bar{\mathrm b}}}

\newcommand{\qq}    {{\rm q\bar{\rm q}}}

\newcommand{\gmgm}    {\gamma \gamma}
\newcommand{\egm}     {{\rm e} \gamma}
\newcommand{\emi}     {{\rm e}^-}

\newcommand{\Mh}      {m_{\mathrm{h}}}

\newcommand{\rts}     {\sqrt{s}}
\newcommand{\rtsee}   {\sqrt{s}_{\rm ee}}
\newcommand{\charpm}  {{\tilde{\chi}}^\pm}
\newcommand{\charp}   {{\tilde{\chi}}^+}
\newcommand{\charm}   {{\tilde{\chi}}^-}
\newcommand{\neut}    {{\tilde{\chi}}^0}
\newcommand{\sel}{\tilde{{\rm e}}_{\rm R}}
\newcommand{\smur}{\tilde{\mu}_{\rm R}}
\newcommand{\smul}{\tilde{\mu}_{\rm L}}
\newcommand{\stau}{\tilde{{\tau}_{1}}}

\begin{document}



\begin{frontmatter}

\begin{flushright}
DESY 06-007\\
26th January 2006
\end{flushright}
\title{
Studies for a Photon Collider at the ILC
}
\author[desy]{F.~Bechtel}
\author[desy,karl]{G.~Kl\"amke}
\author[desy,mbi]{G. Klemz}
\author[desy]{K.~M\"onig\corauthref{cor1}}
\author[desy]{H.~Nieto}
\author[desy]{H.~Nowak}
\author[desy,tem]{A.~Rosca}
\author[desy,flor]{J.~Sekaric}
\author[desy,rwth]{A.~Stahl}

\address[desy]{DESY, Zeuthen, Germany}
\address[karl]{now at Inst.~f\"ur Theo.~Phys., Uni.~Karlsruhe, Germany}
\address[mbi]{now at Max Born Inst., Berlin, Germany}
\address[tem]{now at West Uni.~of Timisoara, Romania}
\address[flor]{now at Florida State Univ., Tallahassee, USA}
\address[rwth]{now at III.~Phys.~Inst., RWTH Aachen, Germany}

\corauth[cor1]{Corresponding author, email: Klaus.Moenig@desy.de}
   
\begin{abstract}
\noindent
One option at the International Linear Collider is to convert the
electron beams into high energy photon beams by Compton scattering a
few millimetres in front of the interaction region. Selected physics
channels for this option have been analysed and technical issues have
been studied. So far no showstoppers for this option have been found.
\end{abstract}
\end{frontmatter}
%
%

\section{Introduction}
\label{sec:intro}
Collisions between photons are interesting in many respects. Up to now they
could only be realised in $\ee$ colliders as collisions between virtual or
quasi-real photons radiated off the electrons. This had the big disadvantage
that the photon energy on average is much lower than the energy of the
electrons. 

Contrary to a storage ring, in an $\ee$ linear collider the beams collide only
once. It should thus be possible to convert the electrons into high energy
photons by scattering them on a high power laser a few mm in front of the
interaction point \cite{Ginzburg:1981ik}. If the laser wavelength is chosen
correctly photon energies up to 80\% of the electron beam energy can be
achieved. 

There are several reasons why a photon collider is interesting for particle
physics \cite{tdr_gg}.  Photons couple equally to all charged particles. Since
there are no interferences involved in the process and the particles are
produced via t-channel exchange the cross sections are in most cases
significantly larger than in $\ee$.  In most cases these cross sections are
just given by the charge of the produced particles and phase space factors. 
This makes the production less interesting than in $\ee$, where the
non-trivial weak couplings of the particles can be measured. However,
the large and model independent cross section offers an excellent possibility
to study the decays of the produced particles.  In the case of W-production
their gauge couplings can be measured, where $\gmgm \rightarrow \WW$ is only
sensitive to the photon couplings of the W without uncertainties from the Z.
Neutral particles can couple to photons only via loops.  The particles running
in the loops can be too heavy to be produced at the collider, but their
effect can be measured from the production cross section of a neutral
particle.  The production of Higgs bosons is very interesting in this respect,
since the coupling of the Higgses to photons is sensitive to all charged
particles that receive their mass from the Higgs mechanism and for example in
SUSY models also to their superpartners.

In addition to the $\gmgm$ mode a photon collider can also be used as an
$\egm$ collider. There is a significant $\egm$ luminosity in the $\gmgm$ mode
due to unconverted electrons, denoted as parasitic $\egm$ mode in the
following. If needed, also only one beam can be converted to get a larger
$\egm$ luminosity with less background.

The detailed layout of a photon collider depends on the parameters of the
linear accelerator.  Especially the time structure of the beam influences
directly the layout of the laser system.  In this paper the layout of the
TESLA machine \cite{tdr_machine} will be used.  The design of the ILC is not
yet finalised, but the time structure will be very similar to the ones studied
for TESLA \cite{ilcbcd}.  A principle layout of a photon collider at TESLA is
presented in \cite{tdr_gg}.

\subsection{Principles of a Photon Collider}

\subsubsection{Compton scattering}

The high energy photons at a photon collider are produced by Compton
scattering of a high energy electron beam with a high power laser.
From pure kinematics the maximum photon energy $\omega_m$ is given by
\[
\omega_m=\frac{x}{x+1}E_0
\]
with $E_0$ being the beam energy and 
\[
x=\frac{4E_0 \omega_0 }{m_e^2c^4}\cos^2{\frac{\alpha_l}{2}}
 \simeq 
 19\left[\frac{E_0}{\TeV}\right]
\left[\frac{\mu m}{\lambda}\right].
\]
$\omega_0$ and $\lambda$ denote the photon energy and wavelength of the laser
and
$\alpha_l$ denotes the crossing angle between the laser and the electron beam.
In the approximation the assumption $\cos \alpha_l/2 = 1$ is used which is
always fulfilled in praxis.
It is desirable to keep $x$ below 4.8, since for larger $x$ the
invariant mass between a high energy photon and a laser photon is above
the $\ee$ pair production threshold, so that the high energy photons are lost
again due their interaction with the laser. The most powerful solid state
lasers have a wavelength around $\lambda \approx 1 \micron$ resulting in 
$x=4.75$ for a 250\,GeV beam.

\begin{figure}[htb]
\centering
\includegraphics[width=0.47\linewidth]{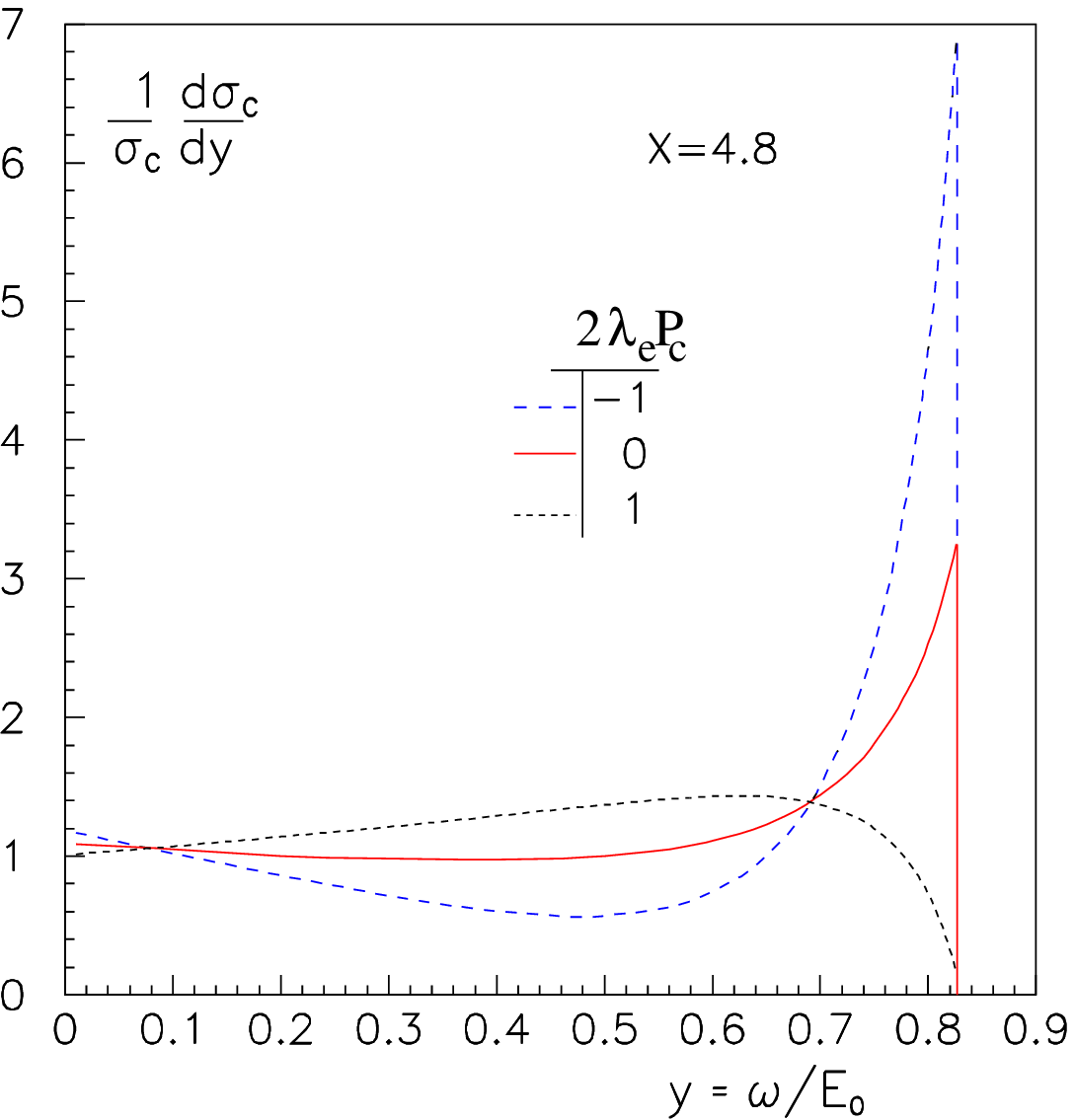}
\includegraphics[width=0.49\linewidth]{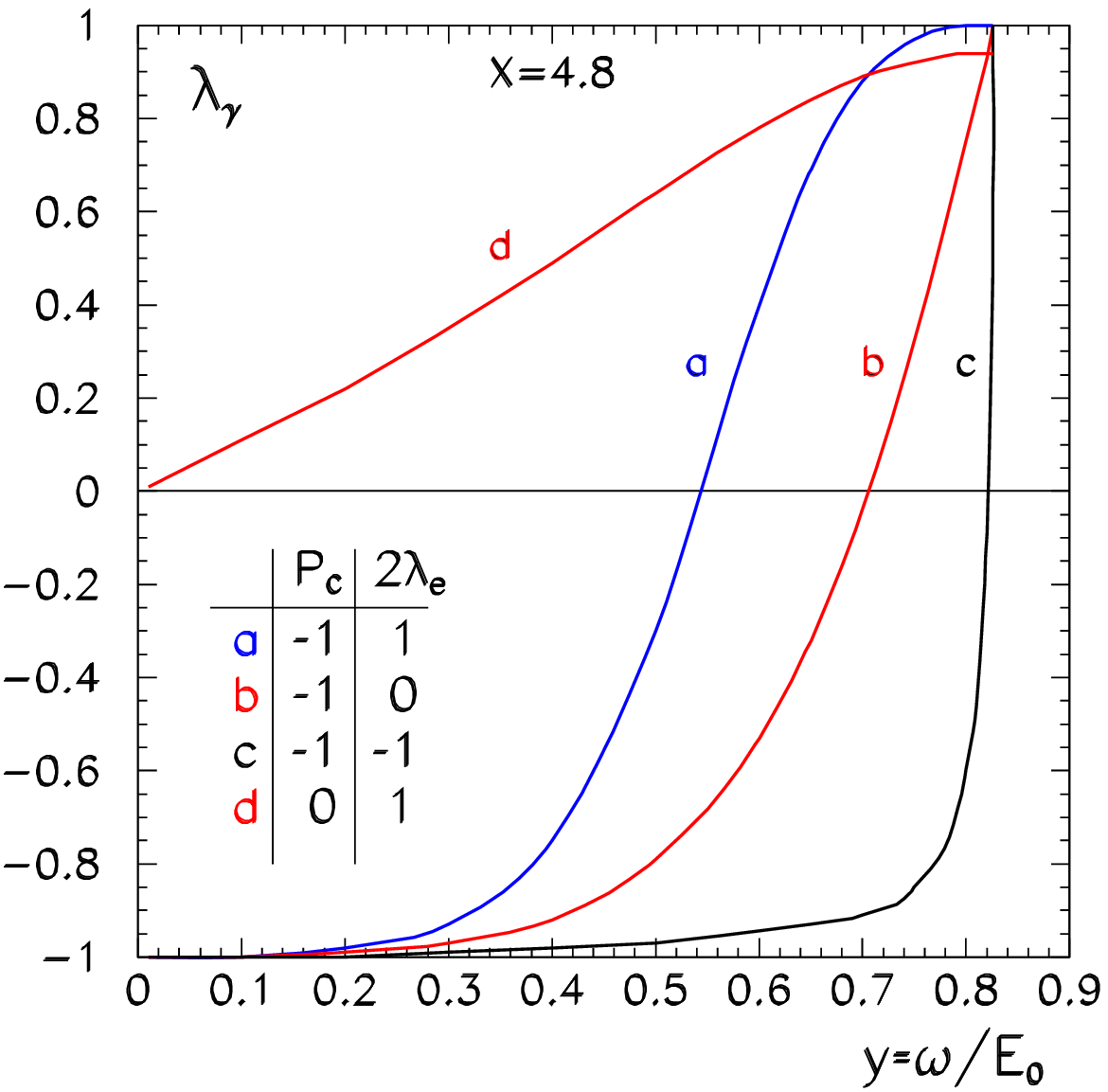}
\caption{Left: normalised photon energy energy spectrum from Compton
  scattering for different electron and laser polarisation; 
  right: photon circular polarisation for different electron and laser
  polarisation.Because of parity conservation in Compton scattering
  the missing combinations can be obtained from the shown ones by
  flipping all involved helicities.} 
\label{fig:intro_compton}
\end{figure}

The differential cross section with respect to the photon energy $\omega_0$
for Compton scattering is given by \cite{tdr_gg}:
\[
{d\sigma _{c}\over d y}   = {2\sigma_{0}\over x}
\left[{1\over 1-y} + 1-y - 4r(1-r) +
2\lambda_{e}P_c r x(1-2r)(2-y)\right]
\]
with $y = \omega_0 /E_0,  r=\frac{y}{(1-y)x}$ and $\sigma_{0}=\pi r_e^2 $.
The cross section is strongly sensitive to the product of the electron
helicity $\lambda_e$ and the laser circular polarisation $P_c$.
The left plot of figure \ref{fig:intro_compton} shows the normalised cross
section for different values of $2\lambda_{e}P_c$. 
For most analyses a maximum luminosity at high energies is desired
requiring a value close to $-1$.
The only exception studied so far is the search for heavy MSSM Higgses
\cite{AsnerHA}. If the mass of these particles is unknown it may be
advantageous to search a larger part of the spectrum simultaneously.
For this a value around zero or even $+1$ may be better.

The helicity of the final state photons is given by 
\[
\langle\lambda_{\gamma}\rangle = 
{-P_c(2r-1)[(1-y)^{-1}+1-y]+2\lambda_{e} xr[1+(1-y)(2r-1)^2]
\over (1-y)^{-1}+1-y-4r(1-r)-2\lambda_{e} P_c xr(2-y)(2r-1)}
\]
The right plot of figure \ref{fig:intro_compton} shows the photon helicity for
different values of the electron and laser polarisation. 
$2\lambda_{e}P_c=-1$ results in a high and relatively constant polarisation in
the high energy peak with a strong variation at lower energies.
For  $2\lambda_{e}P_c=0$ polarising the electron beam is the preferred option. 

\begin{figure}[htb]
\centering
\includegraphics[width=0.6\linewidth]{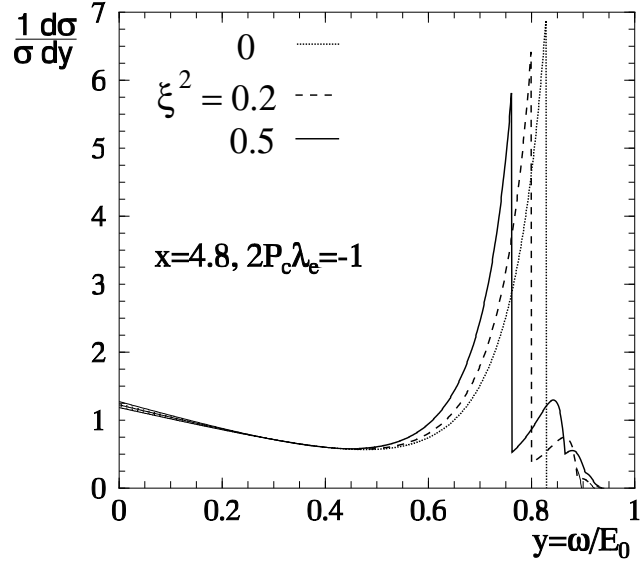}
\caption[]{Normalised photon energy energy spectrum from Compton
  scattering for different nonlinearity parameters \cite{tdr_gg}.}
\label{fig:intro_nonlin}
\end{figure}

The photon spectrum in a real collider is more complicated than the one given
from pure Compton scattering for two reasons. 
The laser power is so high that non-linear effects have to
be taken into account \cite{nonlin1,nonlin2,nonlin3,nonlin4}. 
They can be parametrised by the parameter 
\[
\xi^2 = 
\frac{e^2\bar{F^2}\hbar^2}{m^2c^2\omega_0^2} = 
\frac{2 n_{\gamma} r_e^2 \lambda}{\alpha}
\]
where $\bar{F}$ denotes the field strength of the laser field and $n_\gamma$
the corresponding photon density. The non-linear effects modify the photon
spectrum in two ways. The effective electron mass increases by a factor 
$(1+\xi^2)$ reducing the maximum energy to $\omega_m/E_0 = x/(1+x+\xi^2)$.
At the same time a high energy tail develops due to the simultaneous
interaction of one electron with several laser photons. Figure
\ref{fig:intro_nonlin} shows the photon energy spectrum for different $\xi^2$.

In addition the Compton cross section rises
for smaller centre of mass energies. The electrons that have interacted once
have thus a high chance to interact a second time giving rise to a much
enhanced spectrum at low photon energies.

\begin{figure}[htb]
\centering
\includegraphics[width=0.48\linewidth,bb=4 4 491 473]{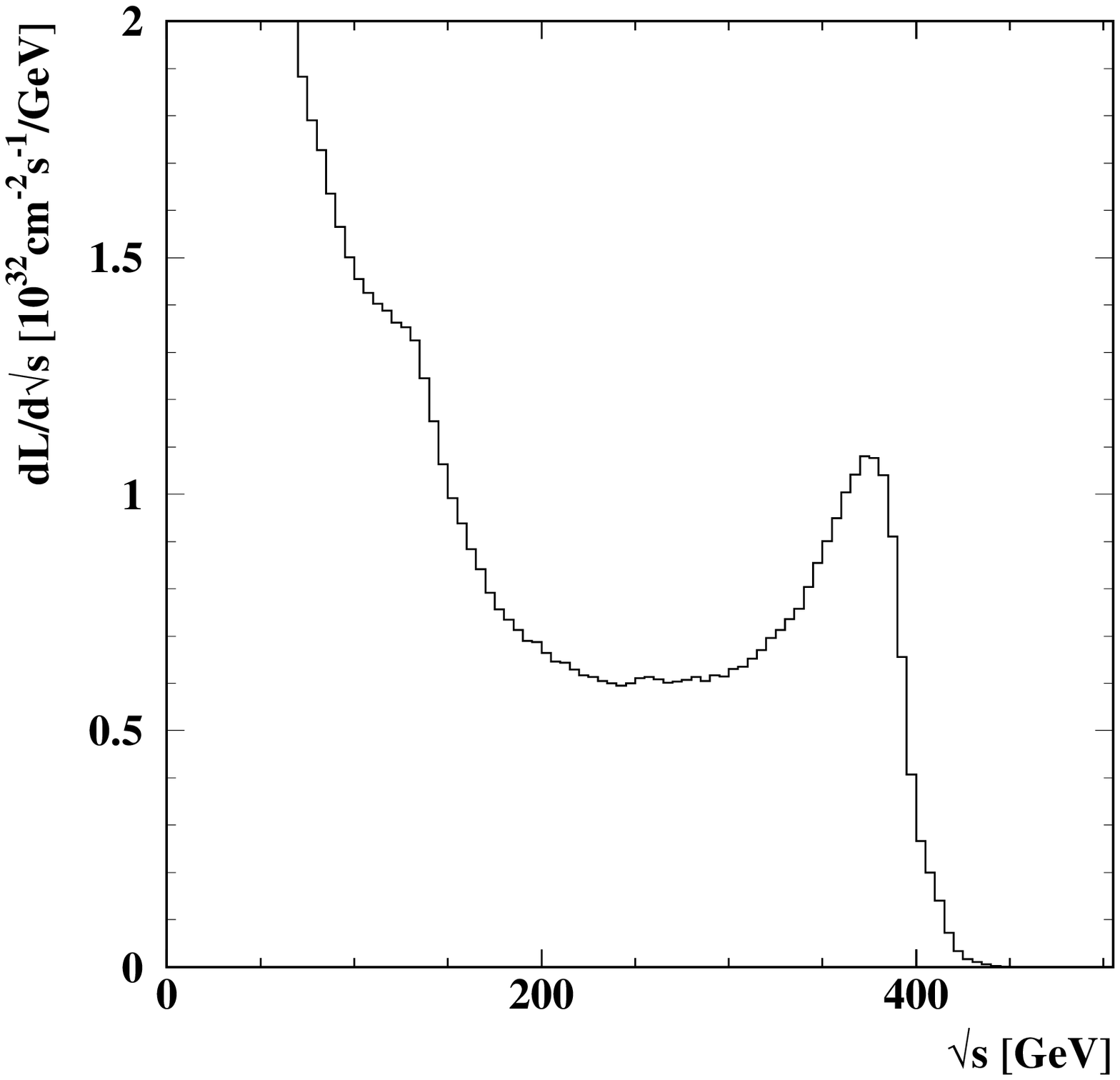}
\includegraphics[width=0.48\linewidth,bb=4 4 491 473]{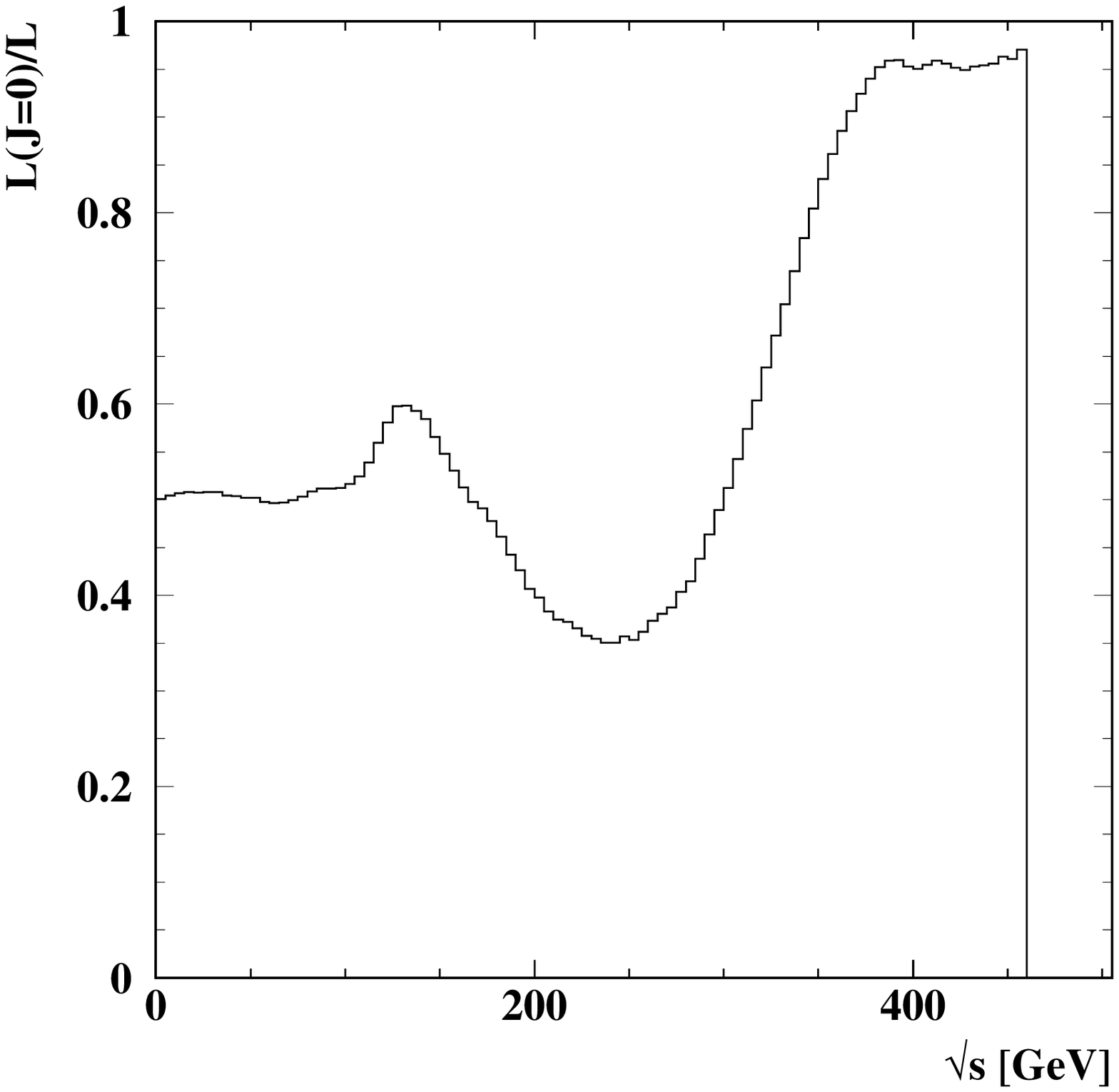}
\caption[]{Luminosity spectrum (left) and fraction of J=0 luminosity
  (right) from CAIN for $\rtsee = 500 \GeV$. The beam parameters labelled
``$\gamma \gamma$ \cite{tdr_gg}'' from table \ref{tab:intro_beam} and the
laser parameters from table \ref{tab:Laser-require} have been used.
}
\label{fig:intro_cain}
\end{figure}

For the calculation of the luminosity spectrum one has to take into
account that lower energy photons are produced at larger angles with respect
to the original beam direction so that their luminous spot is larger, reducing
the luminosity for low centre of mass energies.
To calculate the photon spectrum in a real collider, simulation programs have
to be used. In this paper the program CAIN \cite{cain} is used which
calculates the beam-laser as well as the beam-beam interaction taking
non-linearity and beam polarisation effects into account. Figure
\ref{fig:intro_cain} shows the $\gmgm$ centre of mass energy spectrum and
polarisation for the chosen laser parameters and $\rtsee = 500 \GeV$.

\subsubsection{Machine Parameters}

A photon collider will be operated with two electron beams instead of
an electron and a positron beam for several reasons. Electrons can be
polarised to a higher degree than positrons resulting in a more
favourable luminosity spectrum. Two electron beams defocus each other
in the interaction point generating less beamstrahlung than
electron-positron interactions.  Electron-electron interactions result
in no annihilation events which would be a background for the high
energy $\gmgm$ interactions.  Since the beamstrahlung for identical
beam parameters is smaller at the $\gmgm$ collider than in the $\ee$
case, the beams can be focused stronger resulting in a higher
luminosity.  Table \ref{tab:intro_beam} compares a set of conservative
\cite{tdr_machine} and optimistic \cite{tdr_gg} parameters with the
ones of the $\ee$ collider \cite{tdr_machine} for a beam energy of
250\,GeV.  For the optimistic set a third of the $\ee$ luminosity can
be obtained in the high energy part of the spectrum. Since the
optimistic set results in higher backgrounds it will be used, as a
worst case scenario, consistently in this paper. To reach the same
physics results with the more conservative set the running time has to
be doubled.  The total luminosity with the optimistic design is around
$\mathcal{L} = 10^{35} \lunit$, where about 10\% are in the
interesting high energy region. In the physics studies presented in
section \ref{sec:physics}, it will be assumed that a year of running
corresponds to $10^7$ seconds at design luminosity.

\begin{figure}[htbp]
  \centering
  \includegraphics[width=0.6\linewidth,bb=4 5 567 475]{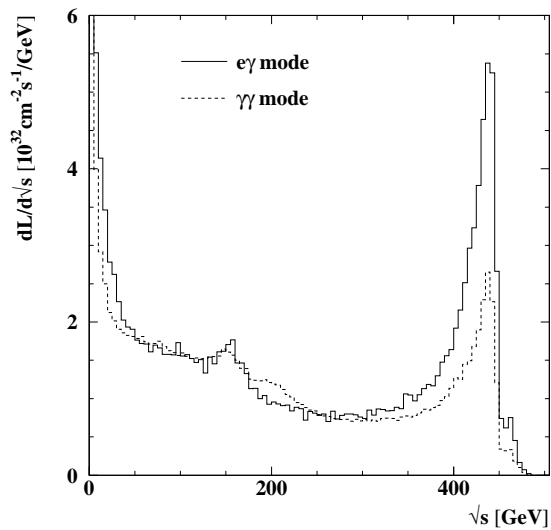}
  \caption{Differential $\egm$ luminosity for the $\egm$- (solid) and 
    $\gmgm$-collider (dashed) at $\rtsee = 500 \GeV$.}
  \label{fig:eglum}
\end{figure}

Figure \ref{fig:eglum} shows the $\egm$ luminosity for the $\gmgm$ and
for the $\egm$ collider at $\rtsee = 500 \GeV$, 
where only one beam is converted and all
other parameters are left identical to the $\gmgm$ collider. The $\egm$
luminosity in the high energy part of the spectrum ($\rts_{\egm} > 360
\GeV$) is $\mathcal{L} = 1.3 \cdot 10^{34}\lunit$ for the $\gmgm$
collider and $\mathcal{L} = 2.4\cdot 10^{34}\lunit$ for the $\egm$
collider.

\begin{table}[htb]
\centering
\caption{Beam parameters for the $\ee$ and the $\gmgm$ collider for $E_b = 250
 \GeV$.} 
\begin{tabular}{|l| c| c| c|} \hline
 & $\ee$ \cite{tdr_machine}& $\gamma \gamma$ \cite{tdr_machine}
 & $\gamma \gamma$ \cite{tdr_gg}\\ \hline
$N/10^{10}$& 2 & 2 & 2  \\  
$\sigma_{z}$ [mm]& 0.3 & 0.3 & 0.3  \\  
pulses/train & 2820 & 2820 & 2820 \\
Repetition rate [Hz] & 5 & 5 & 5 \\
$\gamma \epsilon_{x/y}/10^{-6}$ [m$\cdot$rad] & 10./0.03 &  3./0.03 & 
2.5/0.03 \\
$\beta_{x/y}$ [mm] at IP& 15/0.4 &  4/0.4 & 1.5/0.3 \\
$\sigma_{x/y}$ [nm] & 553/5   & 157/5 & 88/4.3 \\  
$\mathcal{L} (z>0.8z_{m}) $ & 3.4  & 0.6 & 1.1 \\
$[10^{34} {\rm cm^{-2} s^{-1}}]$ & & & \\
\hline
\end{tabular}
\label{tab:intro_beam} 
\end{table}

For parity conserving processes only the total $\gmgm$ and $\egm$
angular momentum matters. The available states are $|J_z| = 0,2$ for
$\gmgm$ and $|J_z| = 1/2,3/2$ for $\egm$\footnote{%
  In the following $J_z$ always denotes the total angular momentum for
  the high energy part of the spectrum.}.
For studies of CP violation
in $\gmgm$ the two $J_z=0$ states $\lambda_\gamma = (1,1)$ and 
$\lambda_\gamma = (-1,-1)$ give different information, while the two
states $J_z = \pm 2$ follow from each other by a rotation of the
coordinate system. Weak processes violate parity, especially the W
couples only to left-handed fermions. For this reason the sign of
$J_z$ is important for many processes in $\egm$.

Because of the large disruption of the electron beam in the interaction with
the laser the spent beam cannot leave the detector through the hole of the
incoming one which is limited by the aperture of the final quadrupole. For
this reason a crossing angle is needed. The exact value depends on the size of
the final focus quadrupole. This study assumes a value of $\alpha =
35\,\mrad$, however with recent quadrupole designs \cite{parker} a smaller
angle seems possible.

\subsubsection{Beam simulation}

For the simulation of the laser-beam and beam-beam interactions many
effects must be taken into account. In the laser-beam interaction
non-linear effects are important where the non-linearity varies with
space and time and particles can interact several times before they
leave the interaction region. In the beam-beam interactions apart
from the interesting high-energy events one has to consider coherent
effects of a particle from one bunch interacting with the coherent field of the
opposing bunch, so called beamstrahlung, and incoherent interactions of
single particles from the two bunches. In addition all processes can
influence the polarisation of the particles.

All processes are included in the simulation program CAIN \cite{cain}
which has been used throughout this paper. Where
possible CAIN has been checked against GuineaPig \cite{guineapig} and
the program of V. Telnov \cite{telsim}. Reasonable agreement has
been found in all cases when a comparison is possible.

For physics studies the beam simulations are not appropriate and
fast programs have been developed which generate the interacting beam
particles and have been tuned to the beam simulation programs.
For the studies presented in this paper the programs CIRCE2
\cite{circe}, based on a histograming technique, and CompAZ \cite{compaz}, 
based on analytic parameterisations, have been used.
%

\boldmath
\section{The Physics Case for a $\gamma \gamma$ Collider}
\begin{figure}[htbp]
\centering
\includegraphics[width=0.8\linewidth,bb=10 17 440 555]{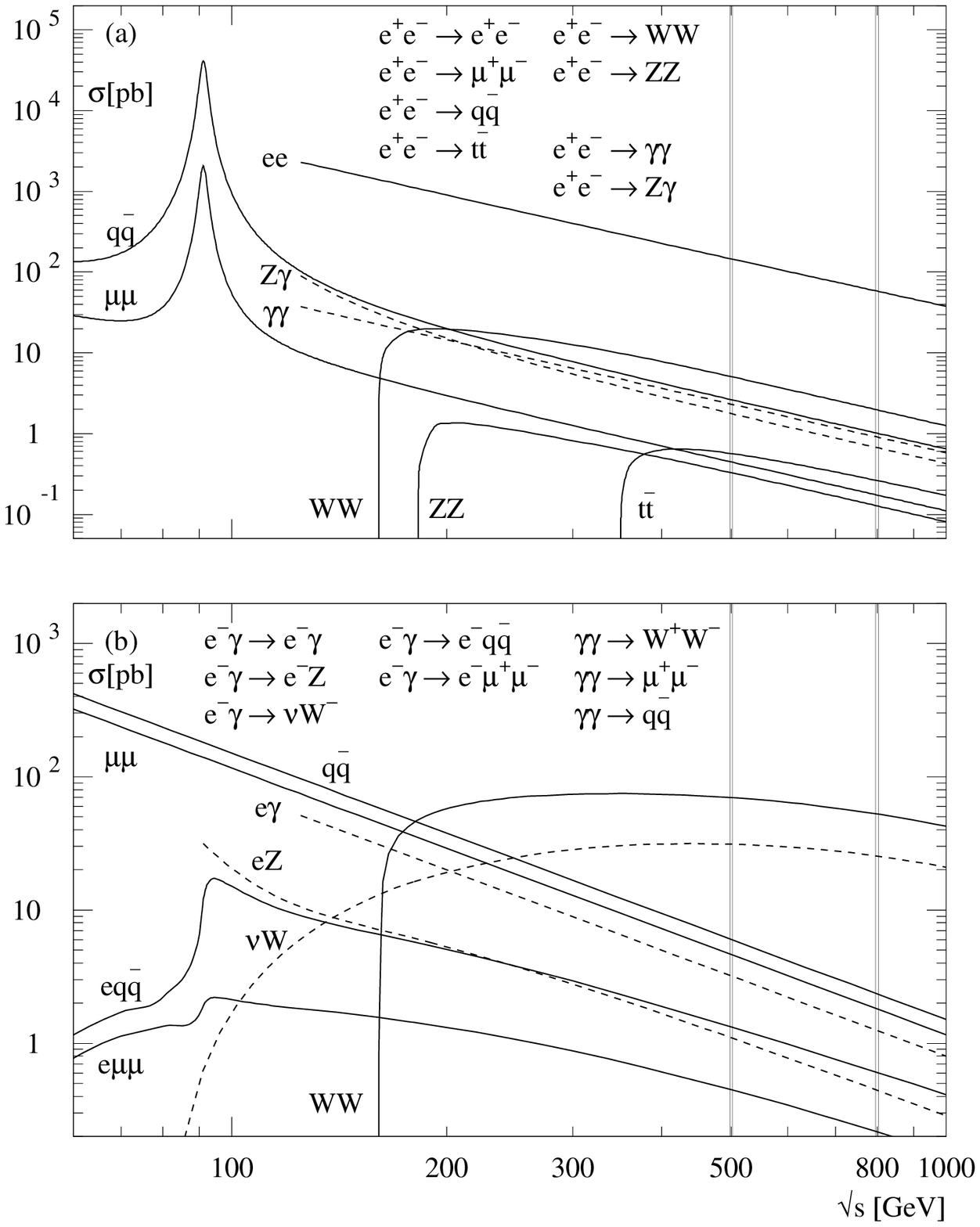}
\caption{Cross section for some processes in $\ee$, $\egm$ and $\gmgm$.}
\label{fig:ggcross}
\end{figure}

\unboldmath
\label{sec:physics}
Photons couple to all charged particles and via loops also to neutral
particles like the Higgs. The pair production of charged particles proceeds
via t-channel exchange where the energy suppression in the propagator is
smaller than for s-channel production. The typical cross sections in $\gmgm$
are thus significantly larger than in $\ee$. The couplings of photons to
charged particles are also well known so that the cross section can be
calculated reliably, contrary to $\ee$, where the production normally proceeds
via photon and Z exchange, where both amplitudes and their relative phase need
to be known. On one hand this makes the production cross section in $\gmgm$ 
much less interesting than in $\ee$. On the other hand the large samples can
be used to learn about the decay properties of the produced particles. 
Figure \ref{fig:ggcross} compares the production cross section of several
particles in $\ee$, $\egm$ and $\gmgm$ \cite{tesla_cdr}.

The physics case of the $\gmgm$ collider is largely complementary to
$\ee$ \cite{tdr_phys}. While in $\gmgm$ the decays can be studied
$\ee$ measured the production mechanism in great detail.

A summary of interesting physics channels can be found in \cite{tdr_gg,gggold}.
In the following a few examples that are of special interest in the motivation
of the $\gmgm$ collider will be discussed.

\subsection{Pileup}
\label{sec:pileup}
The cross section $\gmgm \rightarrow {\rm hadrons}$ is several hundred nb,
relatively independent on the centre of mass energy \cite{pdg}. The total
$\gmgm$ luminosity is $\mathcal{L} = 1.0\cdot 10^{35} {\rm cm^{-2} s^{-1}}$
mainly concentrated at low centre of mass energy (see
Fig.~\ref{fig:intro_cain}), corresponding to several $\mu {\rm b}^{-1}$ per
bunch crossing.  This leads to one to two $\gmgm \rightarrow {\rm hadrons}$
events per bunch crossing, depending on the centre of mass energy, $\rtsee$,
and running mode, that overlay the physics events of interest (pileup).  In
the physics studies these events are taken from a database \cite{hades} and
added to the physics events with the right frequency.  For the 337\,ns bunch
spacing at TESLA the bunch crossing can be tagged unambiguously for every
track so that only single bunch crossings need to be considered.

\begin{figure}[htb]
  \centering
  \includegraphics[width=0.45\linewidth,bb=10 8 495 469]{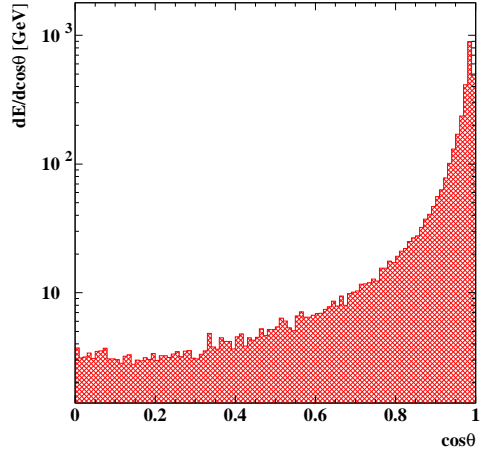}
  \caption{Polar angle distribution for pileup tracks.} 
  \label{fig:pileup_ang}
\end{figure}

Since the $\gmgm \rightarrow {\rm hadrons}$ events are induced by quark
t-channel exchange and have on average a rather large boost in one direction
most tracks are concentrated at low angles (Fig.~\ref{fig:pileup_ang}).
For analyses of channels where most particles are in the central region, like
$\gmgm \rightarrow {\rm H}$ a large part of the pileup can be rejected by a
simple cut on the polar angle.

\begin{figure}[htb]
  \centering
  \includegraphics[width=0.45\linewidth]{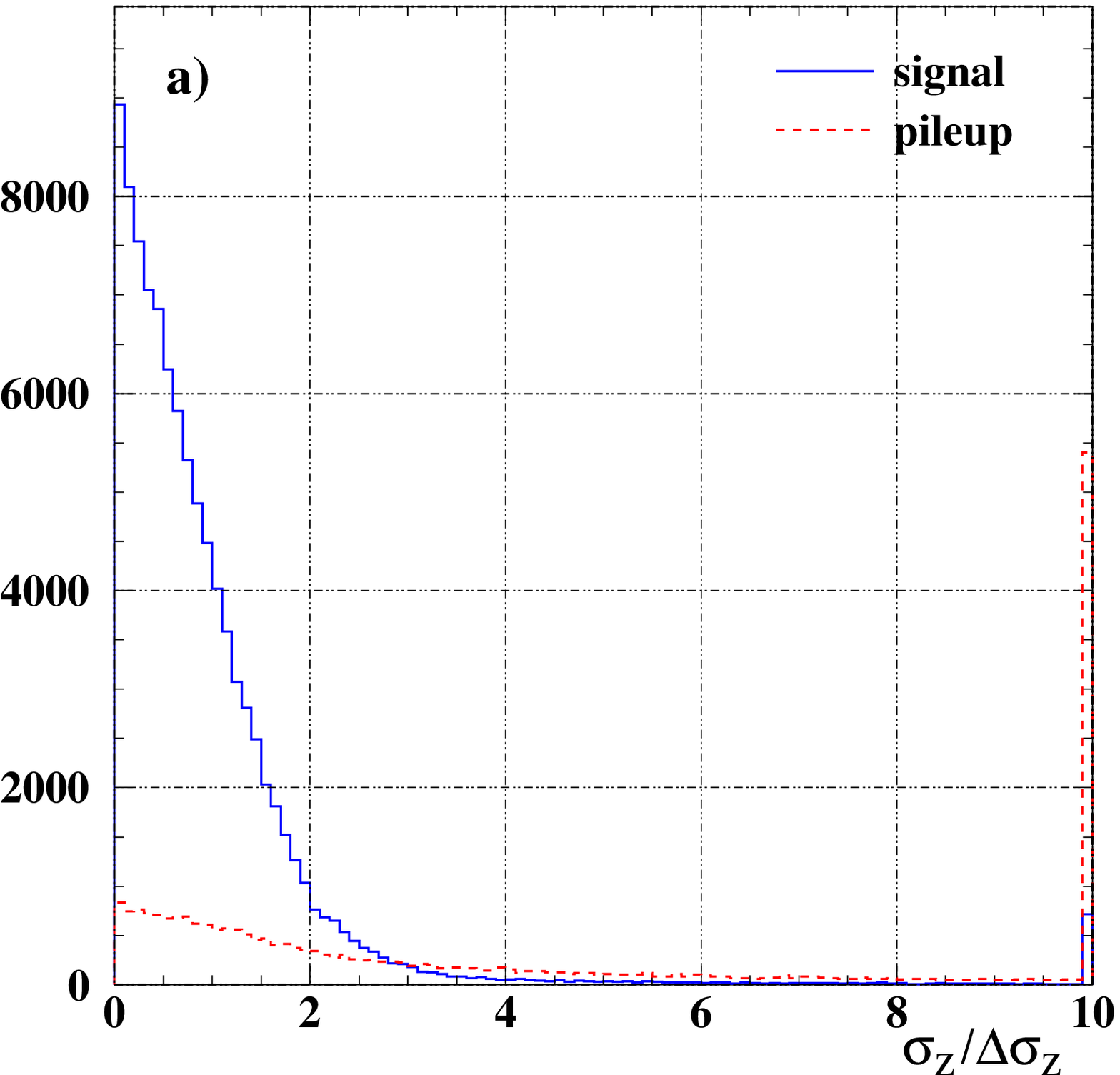}
  \includegraphics[width=0.45\linewidth]{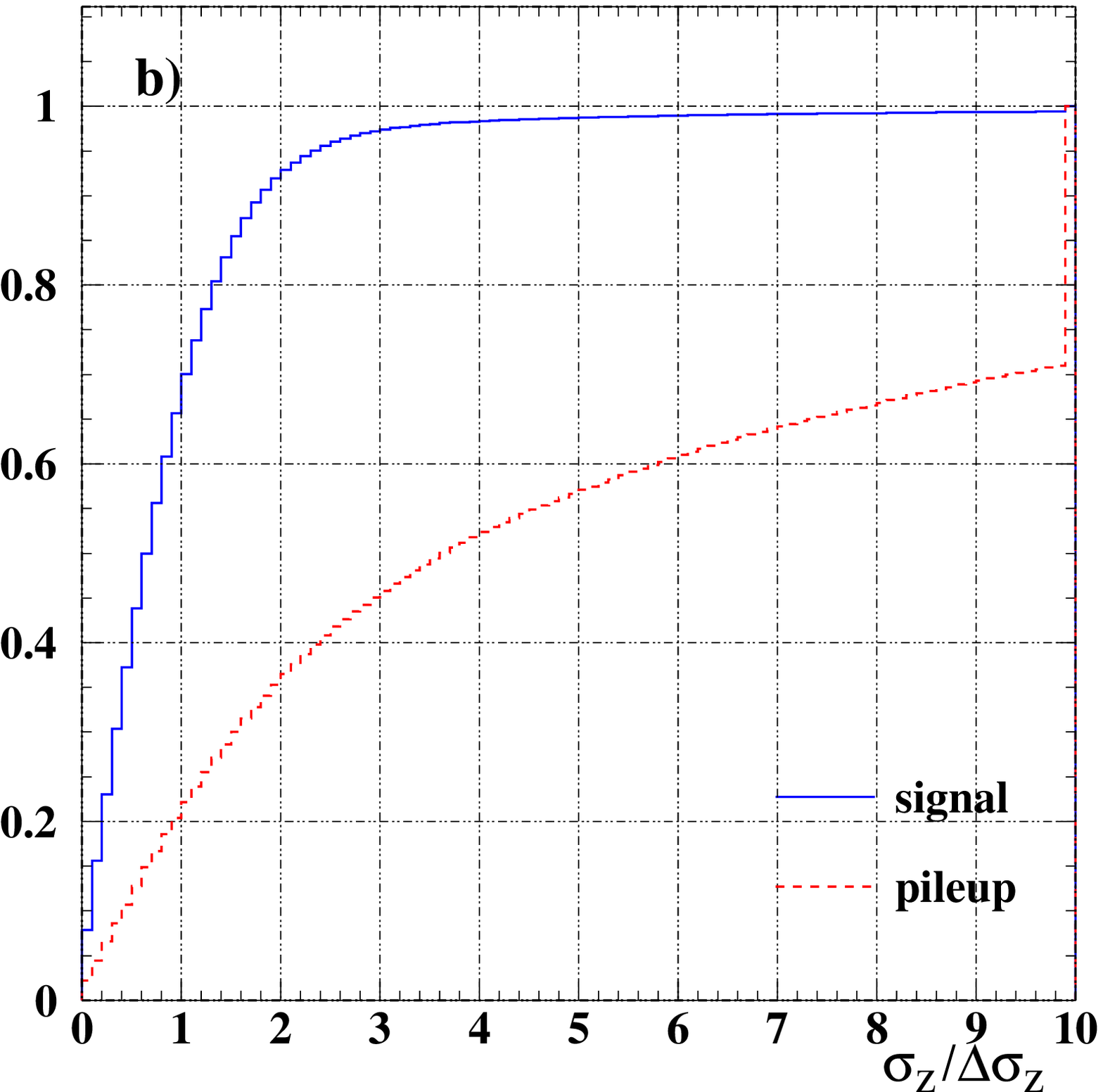}
  \caption{a) Normalised $z$-impact parameter distribution for signal and  
    pileup tracks. 
    b) Track tagging efficiency for a cut on the normalised $z$-impact
    parameter for signal and pileup tracks.}
  \label{fig:imp}
\end{figure}

The bunch length at TESLA is $\sigma_z \approx 300\, \micron$ leading to a
beamspot length of $200\,\micron$. This is much larger than the impact
parameter resolution of the microvertex detector which is around $5\,\micron$
for large momenta.  The measurement of the impact parameter of a particle
along the beam axis with respect to the primary vertex can thus also be used
to separate tracks from the high energy signal and the pileup.
Fig.~\ref{fig:imp} shows the impact parameter in $z$ divided by its error for
tracks from $\gmgm \rightarrow \WW$ and from pileup and the efficiency for a
cut in this variable. For a loss of only 5-10\% of the signal tracks about
half of the tracks from pileup can be rejected.  If enough tracks to
reconstruct the primary vertex are in the central part of the detector and
only tracks in this region need to be considered for the analysis 85\% of the
pileup tracks can be rejected with only a 1\% loss of signal tracks.  This
analysis was done for a channel without b-quarks in the events. In this case
the primary vertex can simply be calculated as the mean momentum weighted
$z$-impact parameter\footnote{The $z$-impact parameter is defined as the
  z-coordinate of the impact point in the $x-y$ plane}. In events with a large
b-quark contents like Higgs production similar results are expected, however a
more sophisticated vertex reconstruction is needed.

For physics, where the tracks
are peaked in the forward region, like the production of
W-bosons, the pileup events are a severe problem. 
In addition the pileup contributes roughly 15 hits
per layer in the microvertex detector. This dominates over the hits from beam
beam interactions in the outer three layers

\subsection{Higgs}
\label{sec:higgs}

In the framework of the Standard Model, the generation of mass of both
fermions and gauge bosons occurs through interactions with the same scalar
particle, the Higgs boson. By the time a Photon Collider is constructed the
Higgs boson, if it exists, will have been discovered. Therefore the aim of
this machine will be a precise measurement of its properties, as for instance
a high precision measurement of the ${\rm H} \to \gamma \gamma$ partial width.
The measurement of this quantity offers an indirect signature for physics
beyond the Standard Model in case a deviation from the Standard Model will be
found.

At a Photon Collider one can measure the product $\Gamma({\rm H} \to 
\gamma \gamma)\times$BR(h $\to$ X). When this
measurement is combined with the measurement of the BR(h $\to$ X) 
from $\ee$ running
one can obtain the partial width independent of the branching ratio.

In this study the process $\gmgm \rightarrow {\rm H} \rightarrow \bb$
has been studied in detail assuming $\Mh = 120 \GeV$.
The feasibility of this
measurement in the intermediate mass region has also been reported by
\cite{higgs1,higgs2,higgs3,higgs4}.

\subsubsection{Simulation of the signal and background processes}

For the Higgs study a total angular momentum of $J_z=0$ is used.
In this case
the cross sections for the direct continuum 
$\gamma \gamma \to \rm b \bar{\rm b}$ and
$\gamma \gamma \to \rm c \bar{\rm c}$ production, the main background
processes, are suppressed by a factor $M_{\rm q}^{2}/s_{\gamma 
\gamma}$, 
with $M_{\rm q}$ being the quark mass. 

The beam spectra at $\rtsee  = 210 \GeV$ are simulated
using the CompAZ \cite{compaz} parameterisation.
The response of the detector has been simulated with
SIMDET4 \cite{simdet}, a parametric Monte Carlo for the TESLA
$\ee$ detector. The only difference between this detector and a
Photon Collider detector is the acceptance of the latter at low polar angles
which is simulated taking for the Higgs reconstruction only
the energy-flow-objects reconstructed 
with a polar angle above 7$^{0}$.

Signal $\gamma \gamma \to {\rm H} \to {\rm b} \bar{\rm b}$ events
corresponding to one year of running are generated using
PYTHIA 6.2 \cite{pythia}. Background processes of the type
$\gamma \gamma \to {\rm q} \bar{\rm q}$(g) are generated with the 
SHERPA \cite{sherpa} generator. A detailed description of the simulation
of the background processes with SHERPA can be found in Reference
\cite{kazimir}.

For $\rtsee = 210 \GeV$ about one pileup event per bunch crossing has
to be taken into account.  A large fraction of this background is
distributed at small angles and it is reduced requiring the cosine of
the polar angle of the energy-flow-objects to be below 0.95.

\subsubsection{Event selection and results}

The selection of the events originating from the $\gamma \gamma \to$h$\to
{\rm b} \bar{\rm b}$ is described in Reference \cite{aura}. Events which
contain two or three jets are selected. Jet are reconstructed using 
the DURHAM \cite{durham} algorithm. The invariant
mass of the jets has to be consistent with the Higgs mass and two
of the jets must be produced by the hadronisation of bottom quarks.

Typical selection variables are: the visible energy of the event; the
longitudinal imbalance in the event, the cosine of the thrust angle 
and the output of the Neural Network \cite{nn} for tagging bottom quarks
for the fastest two energetic jets in the event.

\begin{figure*}[t]
\centering
\includegraphics[width=0.45\linewidth]{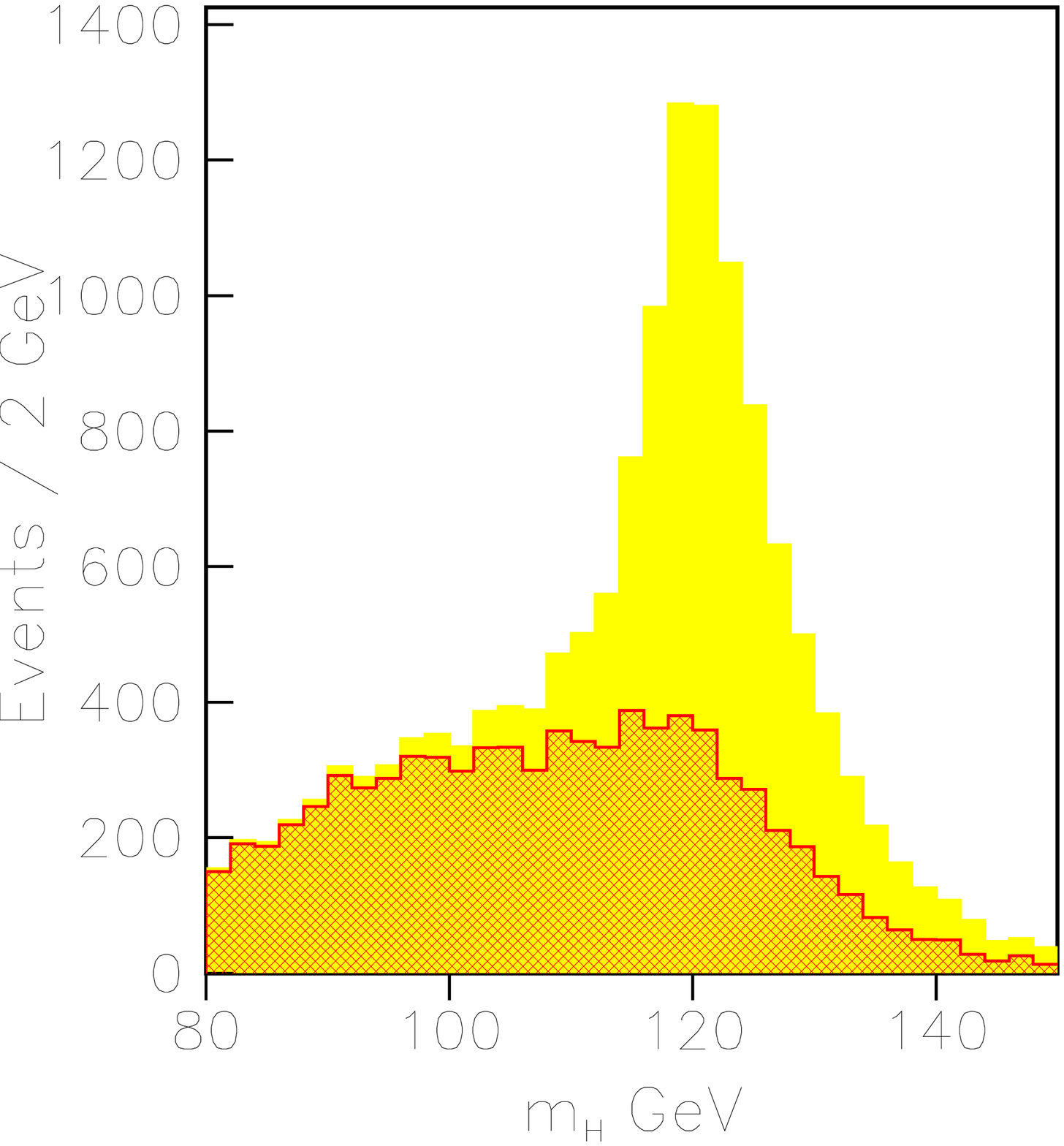} \hfill
\includegraphics[width=0.45\linewidth]{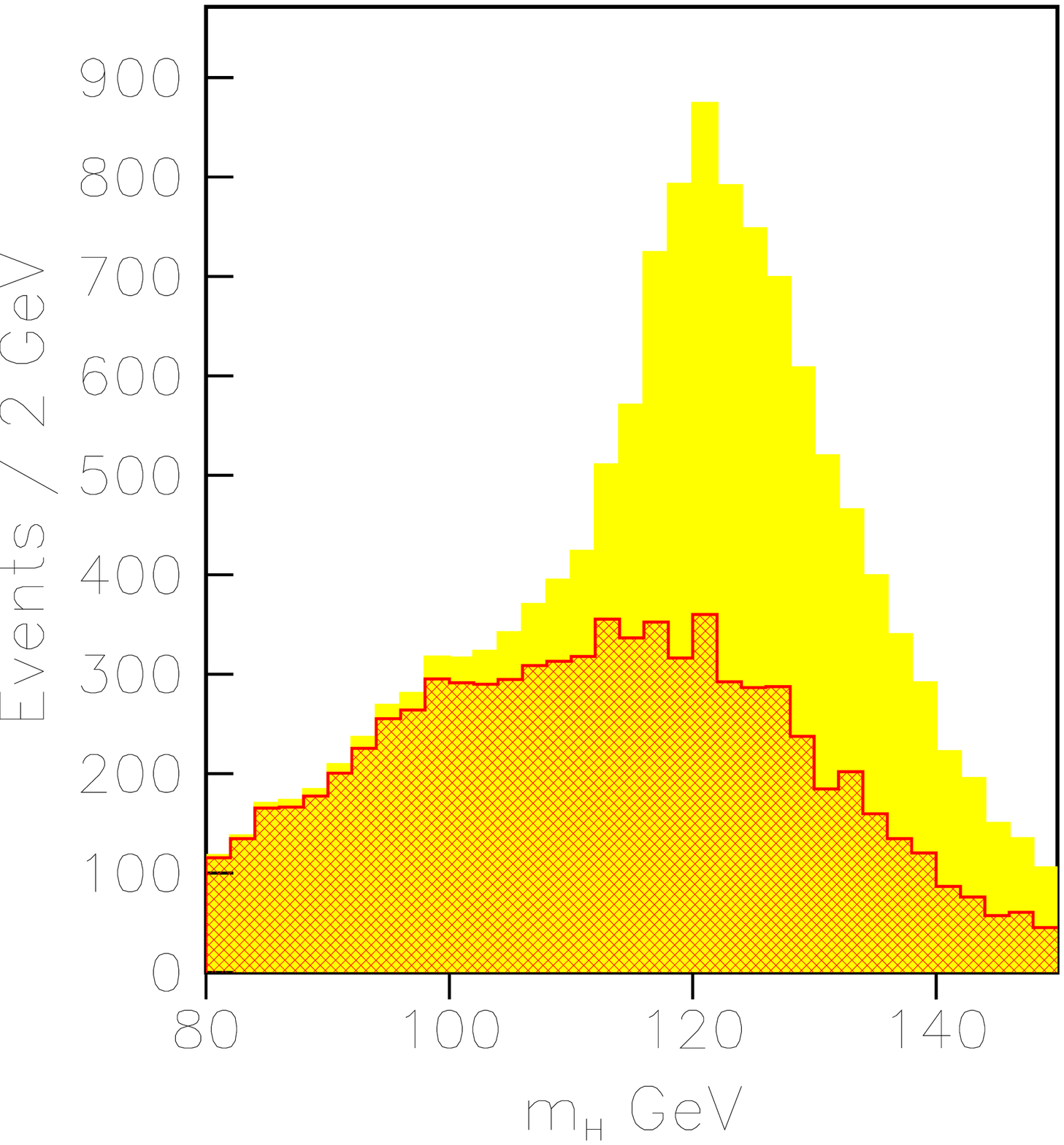} \\
\caption{Distribution of the reconstructed invariant mass
  for the signal and background events without (left) and with (right)
  inclusion of pileup events.}
\label{fig_mass}
\end{figure*}

The reconstructed invariant mass for the selected signal and background
processes is shown in the left part of Figure \ref{fig_mass} if no pileup is
overlaid and in the right part of Figure \ref{fig_mass} when on average one
pileup event is considered per bunch crossing.  To enhance the signal a cut on
the invariant mass is tuned such that the statistical significance of the
signal over background is maximised. Events in the mass region of 116 GeV $ <
M_{jj} < 132 \GeV$ are selected. The number of estimated signal and background
events in this window are 3676 and 2317, respectively.

The two photon decay width of the Higgs boson is
proportional to the event
rates of the Higgs signal. The statistical error of the number of signal 
events,
$\sqrt N_{\rm obs}/({\it N} _{\rm obs}-{\it N}_{\rm bkg})$,
corresponds to the statistical error of this measurement. Here $N_{\rm 
obs}$ is the
number of observed events, while $N_{\rm bkg}$ is the number of expected 
background
events.

\[
\frac{\Delta [\Gamma (\rm H \to \gamma \gamma)\times \rm BR (\rm H \to 
\rm b
\bar{\rm b})]}{[\Gamma (\rm H \to \gamma \gamma) \times \rm BR (\rm H \to 
\rm b
\bar{\rm b})]}=\sqrt N_{\rm obs}/({\it N} _{\rm obs}-{\it N}_{\rm 
bkg})=2.1\% 
\]
can be obtained.
For a Higgs boson with a mass $M_{\rm H}$=120 GeV
the product $\Gamma (\rm H \to
\gamma \gamma) \times \rm BR (\rm H \to \rm b \bar{\rm b})$ can thus
be measured
with an accuracy of 2.1$\%$ using an integrated luminosity corresponding 
to one year of data taking at the TESLA Photon Collider of 
80 fb$^{-1}$ in the high energy part of the spectrum.

\subsection{Gauge Boson Couplings}
\label{sec:tgcs}

\begin{figure}[htbp]
  \centering
  \includegraphics[width=0.35\linewidth]{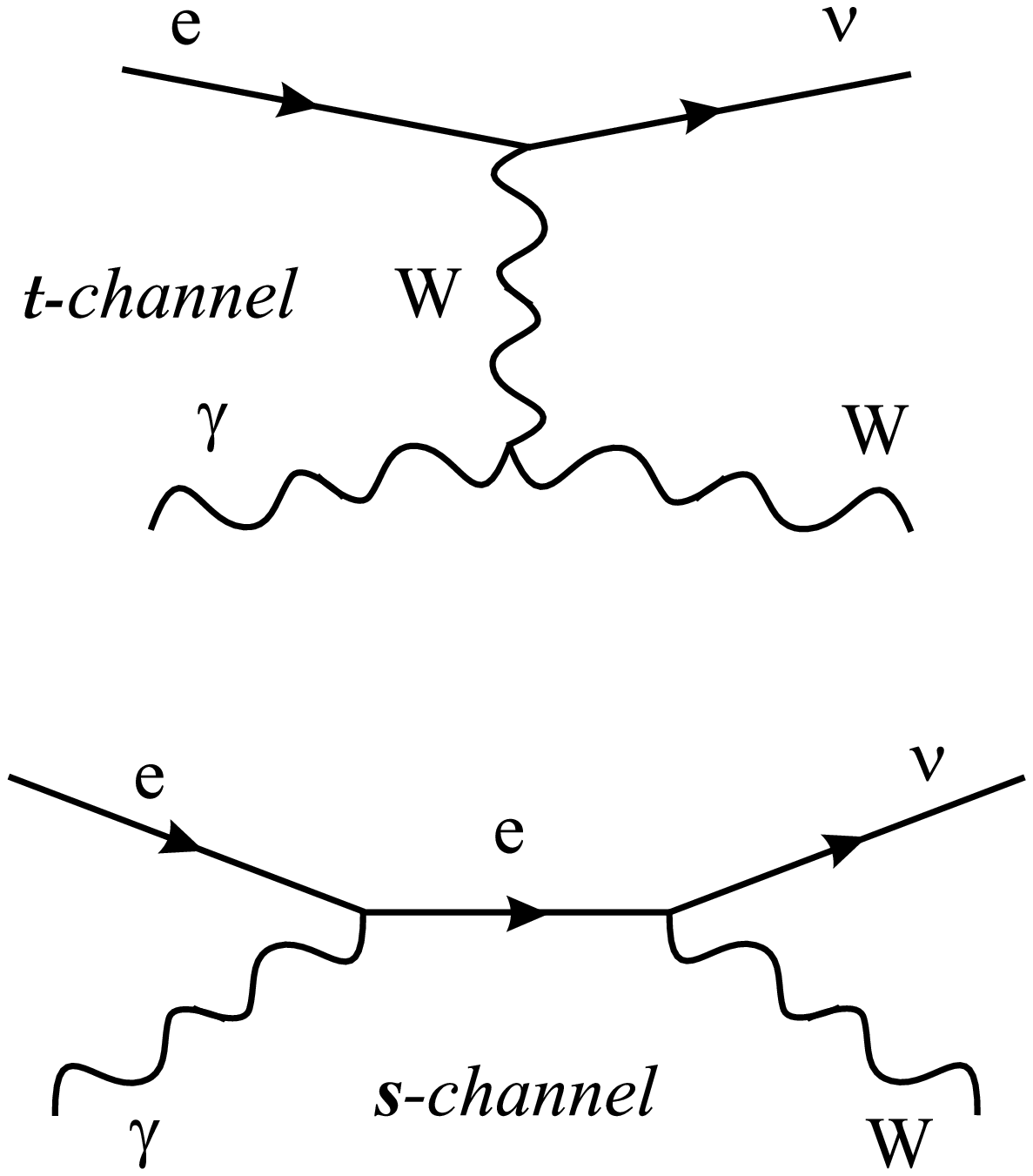}\hspace{1cm}
  \includegraphics[width=0.35\linewidth]{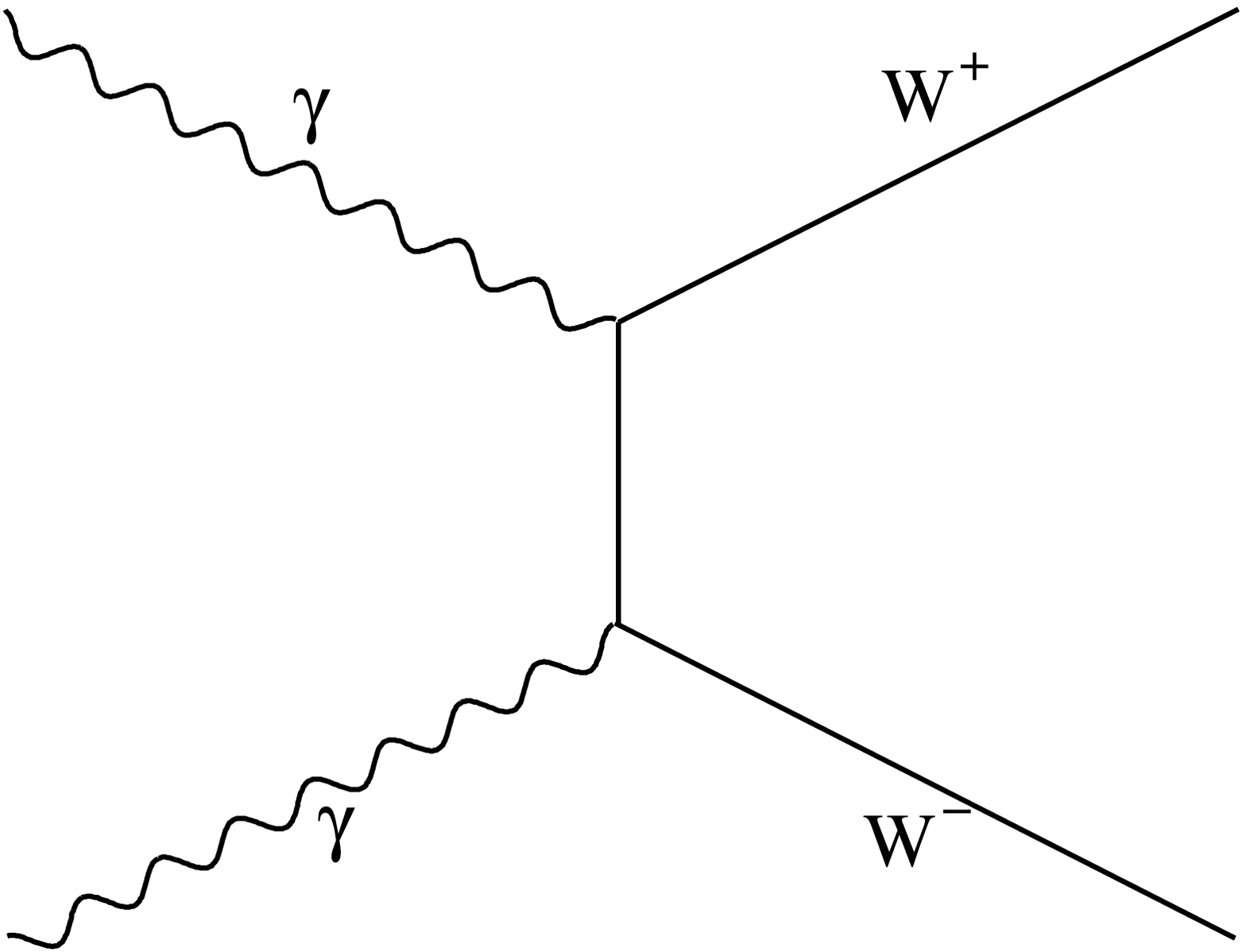}
  \caption{Feynman diagrams for $\egm \rightarrow W \nu$ (left) and $\gmgm
    \rightarrow \WW$ (right).}
  \label{fig:tgcfeyn}
\end{figure}

The triple gauge boson couplings WW$\gamma$ can be measured in the processes
$\egm \rightarrow W \nu$ and $\gmgm \rightarrow \WW$. Both processes have a
large cross section, $\sim 35 \pb$ for $\egm$ and $\sim 80 \pb$ for $\gmgm$
and unpolarised beams. The contributing Feynman diagrams are shown in figure
\ref{fig:tgcfeyn}. In the $\gmgm$ process the only contributing diagram
contains the triple gauge coupling. In the $\egm$ process the electron
t-channel exchange can be switched off if $|J_z| = 3/2$ is used. The cross
sections are about an order of magnitude larger than in $\ee$, however the
sensitivity to the triple gauge couplings is not enhanced by gauge
cancellations.

The C and P conserving couplings $\kappa_\gamma$ and $\lambda_\gamma$
\cite{tgcdef1,tgcdef2} have been analysed in both processes
\cite{jadranka_eg,jadranka_gg}. Since in both cases the beam energy varies and
in addition there is a missing neutrino in $\egm$, a neutrino from a W decay
cannot be reconstructed from the missing four-momentum, only hadronic W-decays
have been used in the analysis. However this is by far not such a large
problem as in $\ee$ since there is no need to separate the $W^+$ from the
$W^-$. Only the sign of the W-helicity cannot be measured in this case.

\subsubsection{Analysis of $\egm \rightarrow W \nu$}

In the $\egm$ case the ``real mode'', where only one electron beam is
converted into photons, and the ``parasitic mode'', where both beams are
converted and the $\egm$ luminosity of the unconverted electrons is used, have
been analysed \cite{jadranka_eg}.  Possible backgrounds in the real mode are
the $\gmgm$ induced process $\egm \rightarrow e \qq$ and $\egm \rightarrow e
Z$ where the electron is lost in the forward region. In the parasitic mode in
addition $\gmgm \rightarrow \qq$ and $\gmgm \rightarrow \WW$ where one W
decays leptonically and the charged lepton is lost, have to be considered.

In the analysis the electron beam always has been assumed to be dominantly
left-handed and the $\egm$ angular momentum is set to $|J_z| = 3/2$, which is
the more sensitive configuration because of the missing s-channel.

\begin{figure}[htbp]
  \centering
  \includegraphics[width=0.4\linewidth]{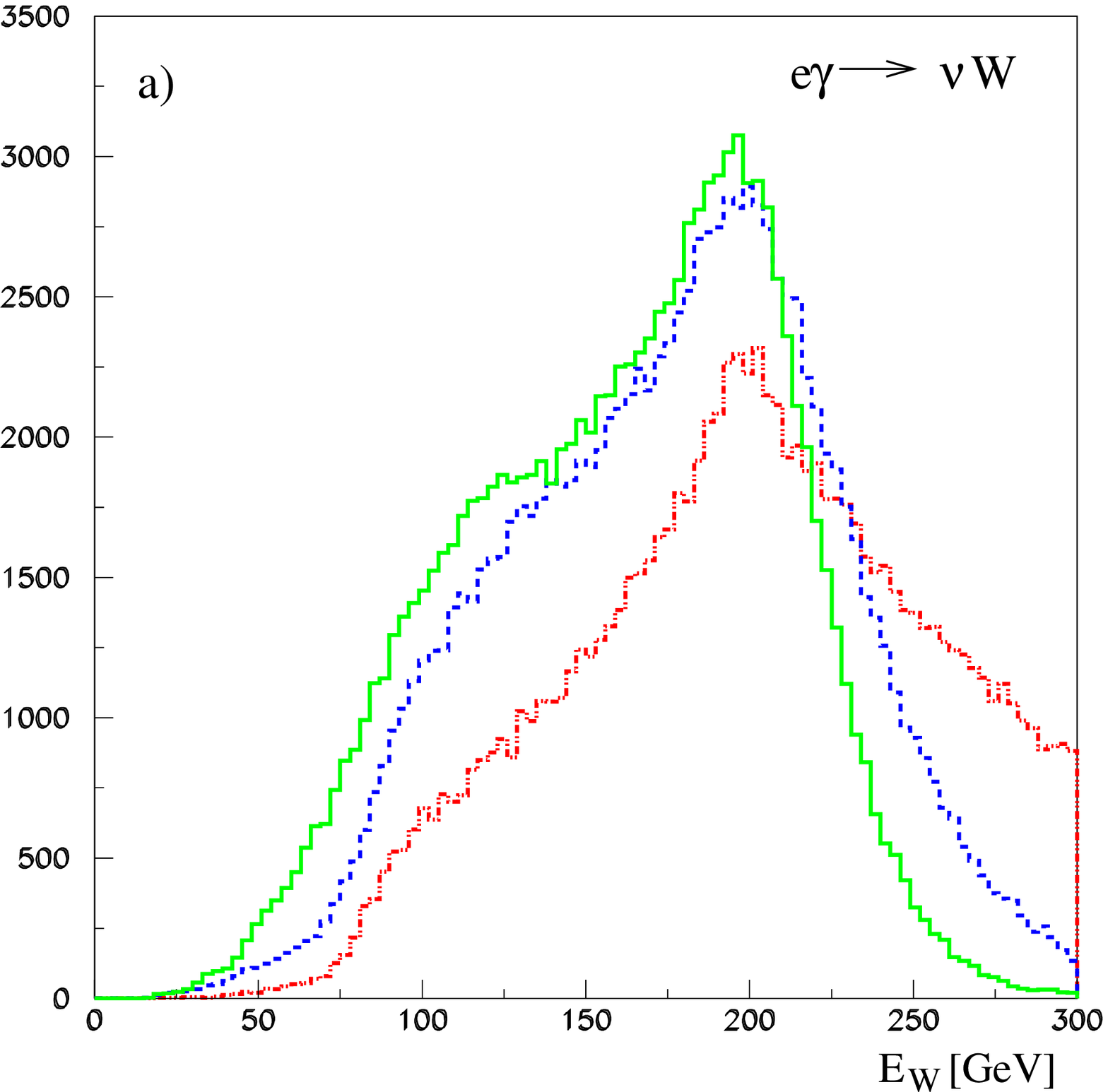}
  \includegraphics[width=0.4\linewidth]{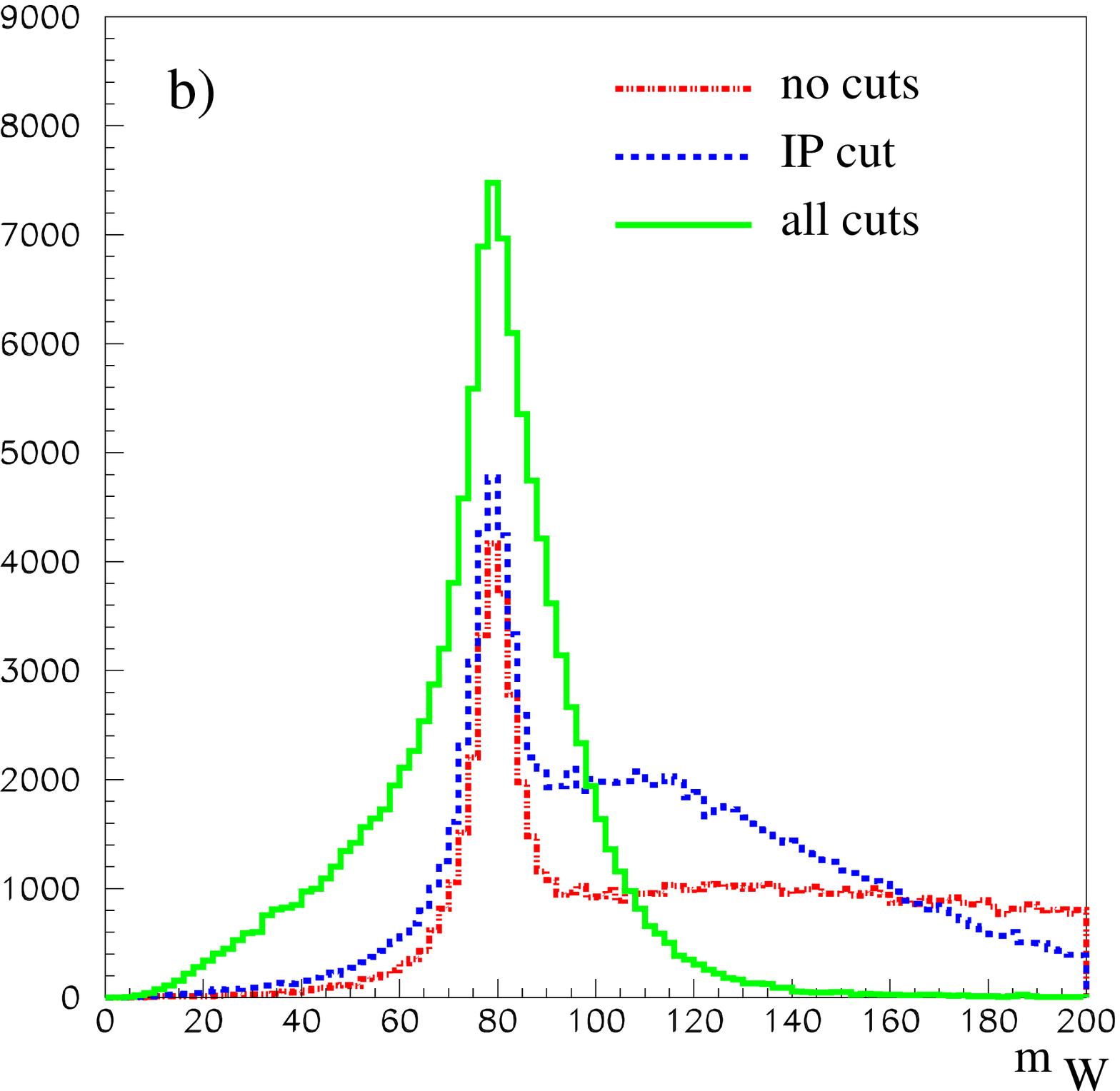}
  \caption{Reconstructed W-energy (a) and -mass (b) at the different
    levels of pileup rejection in the $\egm \rightarrow {\rm W} \nu$
    analysis for the parasitic mode.}
  \label{fig:egpileup}
\end{figure}

In a first step of the analysis pileup tracks are rejected. 
Tracks are rejected if their $z$-impact parameter is inconsistent with the
primary vertex.
In addition
tracks are rejected if they are far away from the reconstructed W-axis where
the cut is stronger if the tracks are in the forward region. Figure
\ref{fig:egpileup} shows the reconstructed W mass and energy distribution
using all tracks, the tracks that pass the impact parameter cut and the tracks
that pass in addition the angle cuts. The large tails get reduced
significantly by the cuts for the price of a worse mass resolution. In a
second step events are selected by a cut on the reconstructed mass and energy
of the W. These cuts result in an efficiency of 73\% (66\%) with a purity of
64\% (49\%) in the real (parasitic) mode. The background is extremely forward
peaked and an additional cut of $\theta_W > 5^\circ$ results in a purity of
95\% (72\%) basically without a loss in efficiency.

The events are fitted using a Monte Carlo reweighting technique where Whizard
\cite{whizard} is used as a generator. In the fit the W production angle, the
two decay angles and the energy are used. In addition to the triple gauge
couplings a free normalisation constant is fitted, where the normalisation is
constrained to unity with an assumed error on the efficiency and luminosity
determination.

Table \ref{tab:egtres} summarises the results for different fit assumptions
assuming one year of running.
The precision on $\kappa_\gamma$ depends mainly in the precision on the total
cross section and thus on the assumed normalisation error. $\lambda_\gamma$
has only little sensitivity to the normalisation but is more sensitive to
pileup and background.
The correlation between $\Delta \kappa_\gamma$ and $\Delta \lambda_\gamma$ is
negligible in all cases.

Several sources of systematic errors have been considered. Assuming
$\Delta \mathcal{L}/\mathcal{L} = 0.1\%$ the beam polarisation needs
to be known to 0.1\% in order that $\Delta \kappa_\gamma$ is not
dominated by this error. At the same time the background has to be
known to a precision of better than 3\% in the real mode and 1\% in
the parasitic mode. $\lambda_\gamma$ is basically not sensitive to
both variations. Both coupling constants are not sensitive to
realistic variations in the luminosity spectrum.

\begin{table}[h]
\caption{
Estimated statistical errors for ${\kappa}_{\gamma}$ and 
${\lambda}_{\gamma}$ from $\egm \rightarrow W \nu$
for the real/parasitic $e\gamma$ mode with different assumptions for one year
of running.}
\begin{center} 
\begin{tabular}{|l|c|c|c|} 
\hline
${{\Delta}\mathcal{L}}$ & 1$\%$ & 0.1$\%$ & 0\\ \hline
 \multicolumn{4}{|c|}{without pileup and background}\\
\hline
${\Delta}{\kappa}_{\gamma}{\cdot}10^{-3}$ & 3.4/4.0 & 1.0/1.0 & 0.5/0.5 \\
${\Delta}{\lambda}_{\gamma}{\cdot}10^{-4}$ & 4.9/5.5 & 4.5/5.2 & 4.5/5.1 \\
\hline
\multicolumn{4}{|c|}{with pileup no background}\\
\hline
${\Delta}{\kappa}_{\gamma}{\cdot}10^{-3}$ &  3.5/4.5 & 1.0/1.0 & 0.5/0.5 \\ 
${\Delta}{\lambda}_{\gamma}{\cdot}10^{-4}$ & 5.2/6.7 & 4.9/6.4 & 4.9/6.4 \\ 
\hline
\multicolumn{4}{|c|}{with pileup and background}\\
\hline
${\Delta}{\kappa}_{\gamma}{\cdot}10^{-3}$ & 3.6/4.8 & 1.0/1.1 & 0.5/0.6 \\ 
${\Delta}{\lambda}_{\gamma}{\cdot}10^{-4}$ & 5.2/7.0 & 4.9/6.7 & 4.9/6.7 \\ 
\hline
\end{tabular}
\end{center}
\label{tab:egtres}
\end{table}

\subsubsection{Analysis of $\gmgm \rightarrow \WW$}

A similar analysis has been performed for $\gmgm \rightarrow \WW$ using both
angular momentum states $|J_z| = 0,2$ \cite{jadranka_gg}. 
The only relevant background for this channel is the process 
$\gmgm \rightarrow \qq$ where four jets are produced from gluon radiation.
Since there are always two W-bosons in the event, typically at low angles,
angular cuts are not effective against pileup, which can thus only be rejected
by the impact parameter cut. Events are selected by cuts on the reconstructed
invariant mass of the W-bosons and the angle between the two jets belonging to
one W in the centre of mass frame of the event. With these cuts an efficiency
of $\sim 50\%$ and a purity of $\sim 80 \%$ has been achieved for both $J_z$
states. 

The data have also been fitted with a Monte Carlo reweighting technique.  In
this case the fit has been done in six dimensions, the W-production angle, two
decay angles per W and the WW centre of mass energy.  Again the two anomalous
couplings have been fitted together with a free but constrained normalisation
factor.  Table \ref{tab:ggres} summarises the results for the two angular
momentum states for one year of running. Like in $e \gamma$ the sensitivity to
$\kappa_\gamma$ is determined by the cross section measurement while
$\lambda_\gamma$ is given by the shape of the events. Consequently
$\lambda_\gamma$ suffers more from the presence of pileup and background.
$\lambda_\gamma$ is determined better from $|J_z| = 2$ while for the same
luminosity error $J_z = 0$ is more sensitive to $\kappa_\gamma$.  However the
luminosity can be determined much more precise in $|J_z| = 2$ so that this
mode is the better one for both coupling constants.

\begin{table}[htb]
\caption{
Estimated statistical errors for ${\kappa}_{\gamma}$ and 
${\lambda}_{\gamma}$ from $\gmgm \rightarrow WW$
for $J_z = 0$/$|J_z| = 2$ with different assumptions for one year
of running.}
\begin{center}
\begin{tabular}{|l|c|c|c|} 
\hline
${{\Delta}\mathcal{L}}$ & 1$\%$ & 0.1$\%$ & 0\\ \hline
 \multicolumn{4}{|c|}{without pileup and background}\\
\hline
${\Delta}{\kappa}_{\gamma}{\cdot}10^{-4}$ & 19.9/29.9 & 5.5/6.2 & 2.6/3.7 \\
${\Delta}{\lambda}_{\gamma}{\cdot}10^{-4}$ & 3.7/3.1 & 3.7/3.1 & 3.7/3.1 \\
\hline
\multicolumn{4}{|c|}{with pileup no background}\\
\hline
${\Delta}{\kappa}_{\gamma}{\cdot}10^{-4}$ & 26.9/37.4 & 5.8/6.8 & 3.0/4.6 \\
${\Delta}{\lambda}_{\gamma}{\cdot}10^{-4}$ & 5.4/4.6 & 5.2/4.6 & 5.2/4.6 \\
\hline
\multicolumn{4}{|c|}{with pileup and background}\\
\hline
${\Delta}{\kappa}_{\gamma}{\cdot}10^{-4}$ & 27.8/37.8 & 5.9/7.0 & 3.1/4.8 \\ \hline
${\Delta}{\lambda}_{\gamma}{\cdot}10^{-4}$ & 5.7/4.8 & 5.6/4.8 & 5.6/4.8 \\ \hline
\hline
\end{tabular}
\end{center}
\label{tab:ggres}
\end{table}

\begin{figure}[htbp]
  \centering
  \includegraphics[width=0.45\linewidth,bb=33 16 494 472]{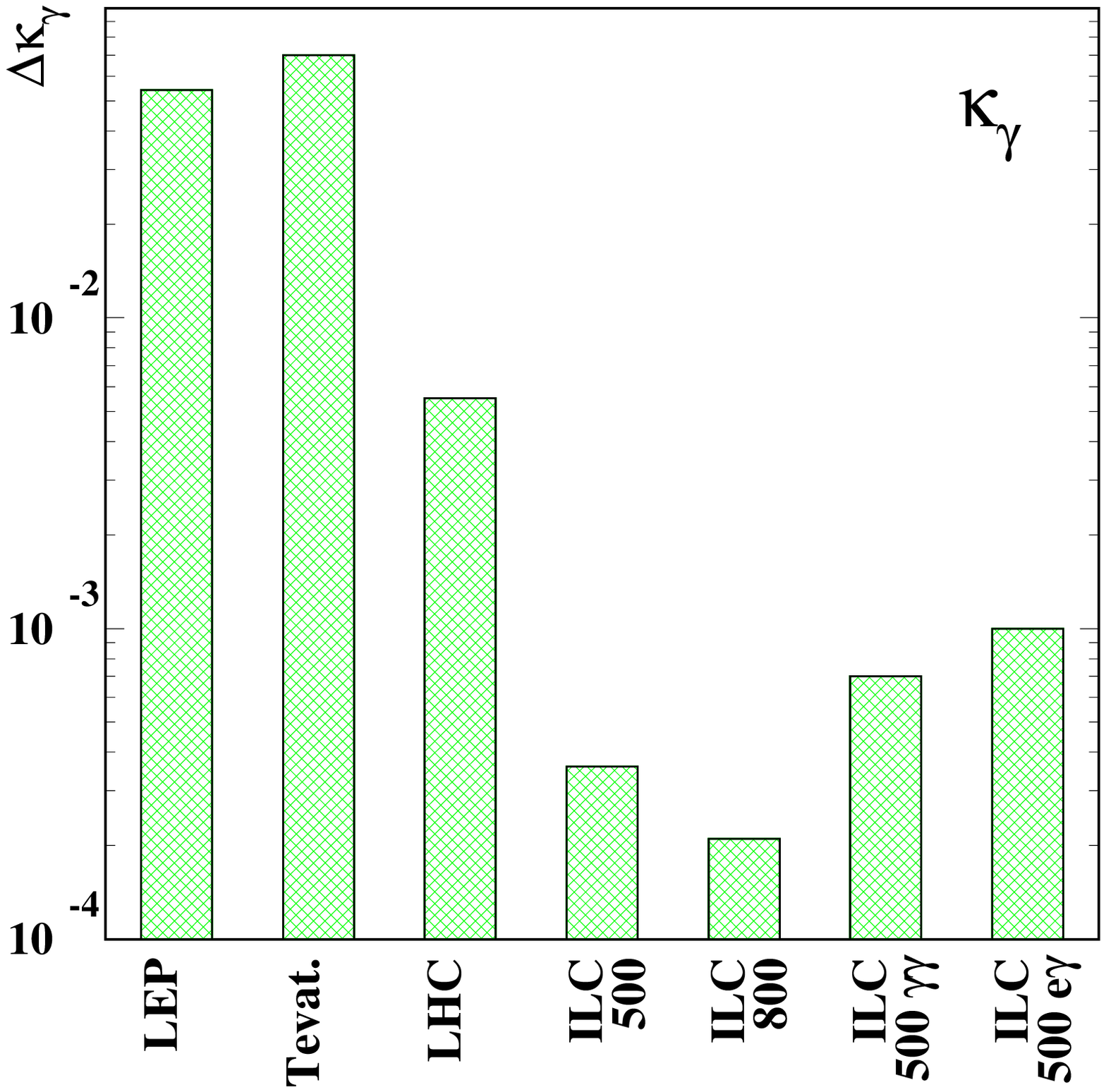}
  \includegraphics[width=0.45\linewidth,bb=33 16 494 472]{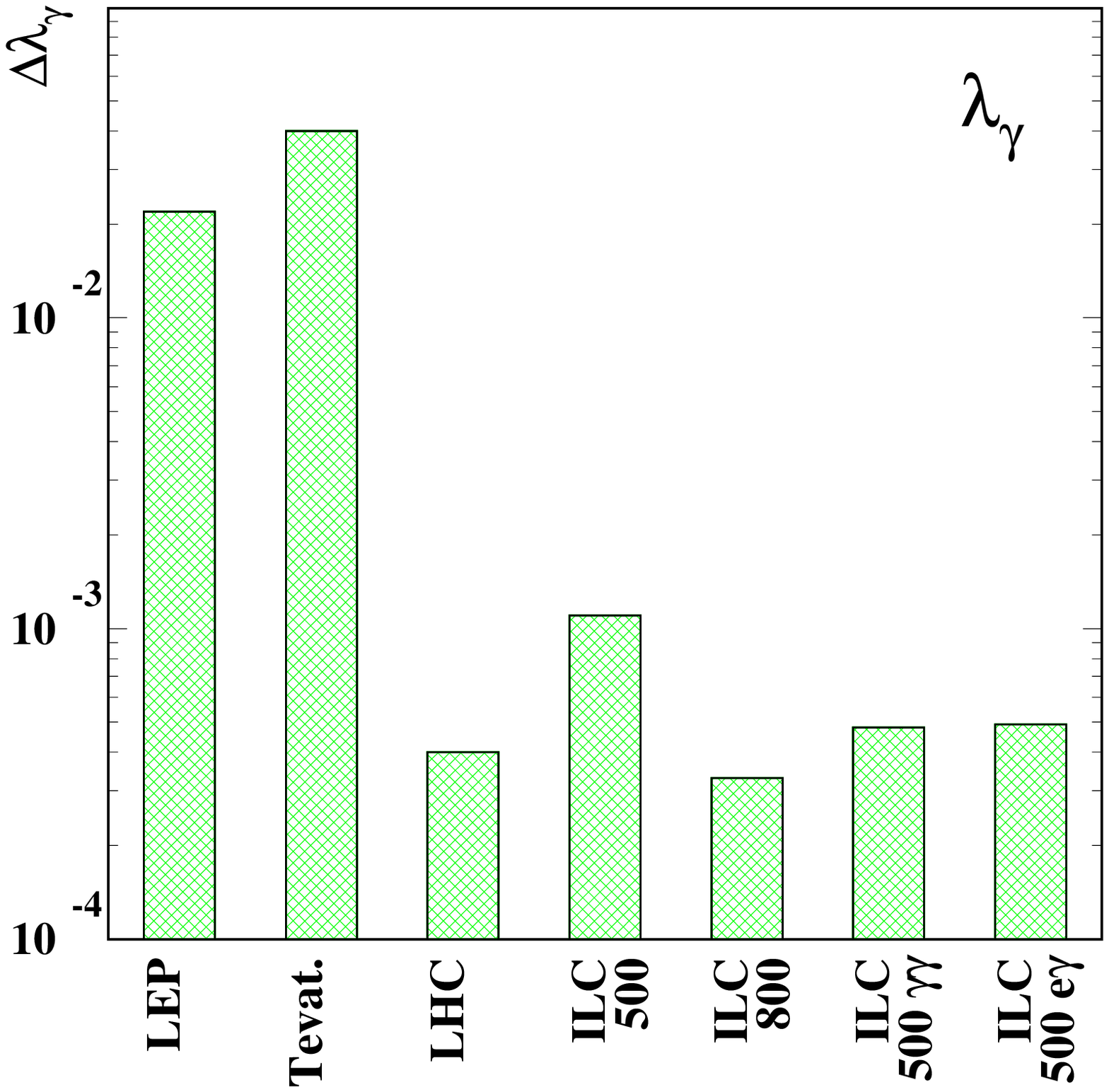}
  \caption{Sensitivity to anomalous WW$\gamma$ couplings at different   
    machines.}
  \label{fig:tgccomp}
\end{figure}

Figure \ref{fig:tgccomp} compares the precision of the fits to $\kappa_\gamma$
and $\lambda_\gamma$ for the different machines. For $\ee$ the results for the
5-parameter fit to $g_Z, \kappa_\gamma, \kappa_Z, \lambda_\gamma, \lambda_Z$
are shown. For $\kappa_\gamma$ $\ee$ is clearly superior to $\gmgm$ and
$\egm$. For $\lambda_\gamma$ the photon collider is better in both modes,
however generator studies indicate that LHC could reach a similar accuracy.

\subsection{Supersymmetry}
\label{sec:susy}

Supersymmetry (SUSY) is generally considered to be the best motivated extension
of the Standard Model \cite{susy1,susy2}. Therefore it will be shown in a few
examples what a photon collider can do for SUSY. Since photons couple simply
to charge the cross sections can be calculated without model dependence and
the photon collider
can thus be used to study decay properties of supersymmetric particles.

As few examples of analyses with supersymmetric particles, sleptons and
charginos in the benchmark point SPS1a \cite{sps} and variants of this
scenario, have been studied. This parameter-point assumes minimal SUGRA with
mass unification at the GUT scale and relatively light superpartners. The
masses and decay modes of the relevant superpartners are shown in table
\ref{tab:sps1}.

\begin{table*}[htb]
  \centering
  \caption{Masses of the Higgses and light superpartners in the SPS1a 
    scenario.}
  \begin{tabular}{|l|c||l|c|}
    \hline
    & $m_{\rm SPS1a}$ &  & $m_{\rm SPS1a}$ \\
    \hline
    \hline
    $h$  & 111.6 & 
    $H$  & 399.6 \\
    $A$  & 399.1 & 
    $H^\pm$ & 407.1 \\
    \hline
    $\neut_1$ & 97.03 &
    $\neut_2$ & 182.9 \\
    $\neut_3$ & 349.2 &
    $\neut_4$ & 370.3 \\
    $\charpm_1$ & 182.3 &
    $\charpm_2$ & 370.6 \\
    \hline
    $\tilde{e}_1$    & 144.9  &
    $\tilde{e}_2$    & 204.2  \\
    $\tilde{\mu}_1$  & 144.9  &
    $\tilde{\mu}_2$  & 204.2  \\
    $\tilde{\tau}_1$ & 135.5  &
    $\tilde{\tau}_2$ & 207.9  \\
    $\tilde{\nu}_e$  & 188.2  & & \\
    \hline
  \end{tabular}
  \label{tab:sps1}
\end{table*}

\subsubsection{Slepton production in $\gmgm$ collisions}

In most models the $\smur$ decays only into $\mu \neut_1$ so
that for $\smur$-pair production probably no information on
Supersymmetry breaking parameters can be obtained.
Nevertheless this process has been simulated inside
SPS1a \cite{huber}. The process is mainly characterised by an
acoplanar muon pair in the detector.  Below the $\smul$ production
threshold the main background is W-pair production. The signal can be
selected with an efficiency of 85\% and a purity of 59\% resulting in
a precision on the cross section measurement of 1.6\% in a month of
running ($100\,\fbi$) with $J_z = 0$. This cross section precision can also be
interpreted as a 0.8\% precision of the branching ratio $BR(\smur
\rightarrow \mu \neut_1)$ where it has to be assumed that possible
other $\smur$ decays don't give any background to the selected
channel.

\begin{figure}[htbp]
  \centering
  \includegraphics[width=0.6\linewidth,bb=4 0 540 685]{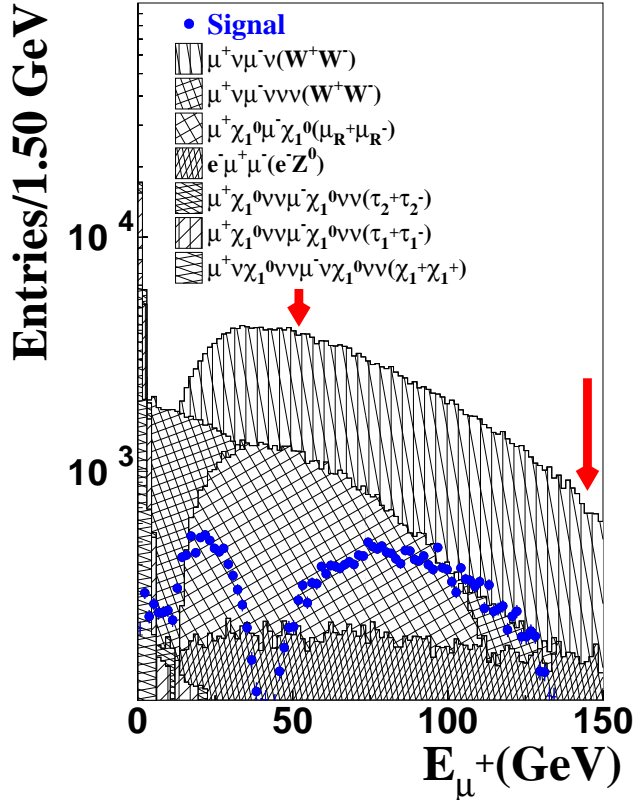}
  \caption{Muon momentum spectrum for $\smul \rightarrow \mu X$
    decays and for the relevant background processes.}
  \label{fig:smulp}
\end{figure}

In the SPS1a scenario the study of $\smul$ decays is much more interesting,
since three decay channels are open $\smul \rightarrow \mu \neut_1$ (55\%),
$\smul \rightarrow \mu \neut_2$ (17\%) and $\smul \rightarrow \nu_\mu
\charpm_1$ (28\%).  $\smul$ pair production has been simulated at $\rtsee =
600 \GeV$ with $J_z = 0$ where one expects $3 \cdot 10^4$ events in one year of
running \cite{huber}. Figure \ref{fig:smulp} shows the muon momentum
distribution from $\smul$ decays together with the expected background. For
the signal one can see three distinct regions. At high momenta there are only
muons from the $\smul \rightarrow \mu \neut_1$ decay, the muons at medium
momenta are from the $\smul \rightarrow \mu \neut_2$ decay and the ones at low
momenta are from $\neut_2$ decays.

Selecting the high momentum region one can with similar cuts as for
the $\smur$ analysis measure the cross section 
$\gmgm \rightarrow \smul++ \smul^- \rightarrow \mu^+ \neut_1 \mu^- \neut_1$
with a precision of around 2\% and the corresponding  branching ratio
${\rm BR}(\smul \rightarrow \mu \neut_1)$  with a precision of around 1\%.
This branching ratio should be sensitive to the mixing angles in the
chargino and neutralino sector.

\subsubsection{Chargino production in $\gmgm$ collisions}

A chargino can decay either into a W and a neutralino or, if kinematically
allowed, into a slepton and the corresponding lepton\footnote{Other decays
  like decays into virtual squarks are usually suppressed}. The flavour
composition of the leptons is non-trivial and depends on the SUSY-breaking
parameters. Within SPS1a the $\charpm_1$ decays almost exclusively into 
$\stau \nu_\tau$,
so that no meaningful branching ratio measurement is possible. Therefore for a
simulation study $m_0$ has been changed to $130 \GeV$ and $\tan \beta$ to 9.
With these parameters the $\charpm_1$ decays in 26\% of the cases into 
$W \neut_1$ and in 73\% into $\stau \nu_\tau$.
The process $\gmgm \rightarrow \charp_1 \charm_1$ has been simulated with
both charginos decaying into ${\rm W} \neut_1$ and the W decaying
hadronically.
Details of the analysis can be found in \cite{gunnar}.
Assuming that the production cross section is
known the event rate is proportional to
${\rm BR}(\charpm_1 \rightarrow {\rm W} \neut_1)^2$.

Because of the huge WW background an efficiency of 24\% and a purity
of 11\% is possible at $\rtsee = 600 \GeV$, leading to a relative
branching ratio error of 3.5\%. To test the usefulness of this
measurement it has been injected into the fit of the low energy SUSY
breaking parameters with Fittino using the masses and cross sections
from the LC/LHC study \cite{fittino}.  Due to the $\charpm_1$
branching ratio measurements the precision on $\tan \beta$ and on the
$\tilde{\tau}$ mixing parameter improve by a factor two to three.
However up to now no decay related observables from $\ee$ are used in
the fit.

\subsubsection{Selectron production in ${\rm e} \gamma \rightarrow \sel \neut_1$}

If the mass difference $\sel \neut_1$ is large it is possible that the
reaction $\ee \rightarrow \sel^+ \sel^-$ is not accessible while ${\rm
  e} \gamma \rightarrow \sel \neut_1$ can be seen at the same $\emi
\emi$ centre of mass energy. In mSUGRA this happens e.g. if $m_0$ and
$m_{1/2}$ are of approximately the same size.  If this happens in
mSUGRA the chargino and neutralino sector would be accessible in $\ee$
via $\ee \rightarrow \neut_1 \neut_2,\, \neut_2 \neut_2,\, \charp_1
\charm_1$ so that $m_{1/2}$ could be measured already there while
$m_0$ can only be obtained from the $e \gamma$ mode. Without gauge
unification it is also possible to adjust the parameters in a way that
no visible SUSY signal is present in $\ee$.

In most models the $\sel$ decays dominantly into ${\rm e} \neut_1$ so that if
R-parity is conserved the experimental signal is a single electron in the
detector. There are two irreducible background channels: 
$\egm \rightarrow {\rm W} \nu$ with ${\rm W} \rightarrow {\rm e} \nu$ and
${\rm W} \rightarrow \tau \nu \rightarrow {\rm e} 3\nu$ and
$\egm \rightarrow {\rm eZ} \rightarrow {\rm e} \nu \bar{\nu}$.
In most cases the lightest sfermion is the partner of the right-handed
electron, so that the signal is enhanced using right-handed electron
beams. This polarisation suppresses simultaneously the W-background which is
present for left-handed electrons only. Also the Z-background is reduced
slightly in this case.

Signal and background have been simulated using SHERPA \cite{sherpa} where the
following SUSY parameters have been used:
\begin{eqnarray*}
m_0 & = & 250 \GeV\\
m_{1/2} & = & 250  \GeV\\
A & = & 0\\
\tan \beta & = & 50 \\
{\rm sign} (\mu) & = & +1
\end{eqnarray*}
resulting in
\begin{eqnarray*}
m(\sel) & = & 273 \GeV\\
m(\neut_1) & = & 100 \GeV\\
m(\neut_2) & = & 190 \GeV\\
m(\charpm_1) & = & 190 \GeV
\end{eqnarray*}
with the $\sel$ decaying almost exclusively into ${\rm e}\neut_1$
\cite{ignacio}.  With cuts in the electron energy and polar angle an
efficiency of $\varepsilon = 71\%$ and a purity $p=63\%$ can be reached
allowing a 1\% cross section measurement in one year of running.  As shown in
\cite{tdr_phys} the selectron and neutralino mass can be obtained from the two
endpoints of the electron energy spectrum in case of monoenergetic beams. For
the simulated parameter set the neutralino mass can be measured in $\ee$
running, so that one endpoint is sufficient to obtain the selectron mass.
Because of the sharp upper edge of the photon energy spectrum the lower
endpoint is approximately preserved. The upper edge gets distorted by the
dependence of the $\sel$-energy from its polar angle but can be reconstructed
from the electron transverse momentum.  It has been estimated that from the
lower endpoint the selectron mass can be measured with a precision of 0.5\%.
The precision from the upper endpoint is significantly worse.

\subsection{Luminosity Measurement}
\label{sec:lumi}
As for any collider, the luminosity at a $\gmgm$ and $\egm$ collider has to be
measured using gauge reactions with a large and well known cross
section. Since the photon polarisation depends strongly on its energy,
the luminosity spectrum has to be measured separately for the
different polarisation states. A general discussion about the
luminosity measurement at $\gmgm$ and $\egm$ colliders can be found in
\cite{lumi_telnov}.

A QED process involving only leptons is the
ideal gauge process since it has basically no unproven physics
assumptions involved.  
The process $\gmgm \rightarrow \ee, \mumu$ has a cross section of a
few pb with realistic tracking cuts for a total $\gmgm$ angular
momentum of $|J_z| = 2$ \cite{lumi_telnov}. This allows for a luminosity
precision around $0.1\%$ in one year of running. For $J_z = 0$ the
cross section is suppressed by $m^2/s$ because of helicity conservation
and thus not usable for luminosity determination.

The process $\gmgm \rightarrow \ell^+ \ell^- \ell^+ \ell^-$
has in principle a very large cross section independent from the beam
polarisation. Unfortunately the leptons in this case are mostly at
very low polar angle where they cannot be measured \cite{lum_4l}.

W-pair production has a very large cross section ($\sim 80 \pb$) for both
polarisation states. However its size depends on the triple gauge
couplings which must be assumed to use the process as a gauge process.
If the gauge couplings are measured in $\ee$ or $\gmgm$ with $|J_z|=2$,
W-pair production can be used for a luminosity measurement 
in $J_z=0$ with a precision better than
1\%. This is largely sufficient for a possible measurement of heavy
SUSY Higgses or superpartners. 

\begin{figure}[htbp]
  \centering
  \includegraphics[width=0.6\linewidth]{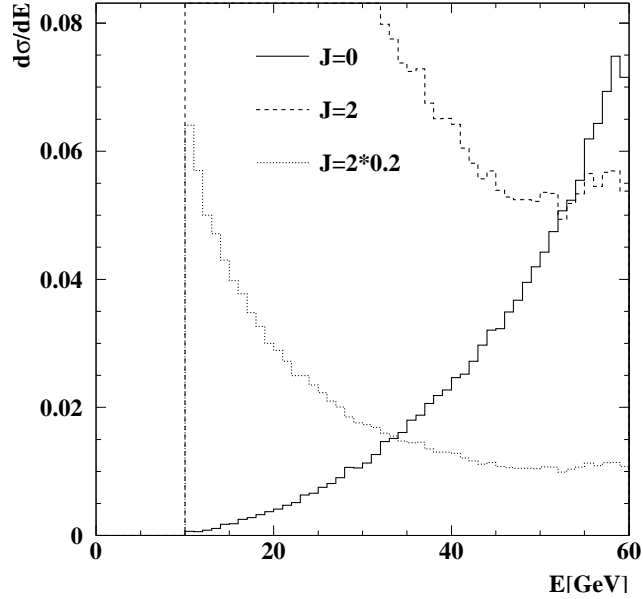}
  \caption{Cross section $\gmgm \rightarrow \ell^+ \ell^- \gamma$ for
    $|J_z|=2$ and $J=0$. The dotted line shows the $|J_z|=2$ cross
    section multiplied by 0.2.}
  \label{fig:llg}
\end{figure}

The accurate measurement of the partial width ${\rm H} \rightarrow
\gmgm$ for a light Standard Model like Higgs of course has to be done
with $J_z = 0$.  Its mass is either below the W-pair production
threshold or the Higgs decays dominantly into a W-pair, so that W-pair
production can never be used as a gauge process.  A candidate gauge
process here is $\gmgm \rightarrow \ell^+ \ell^- \gamma$.  Because of
the extra radiated photon the mass suppression does not apply and the
cross section for monoenergetic beams is around 1\,pb for $\sqrt{s} =
120\GeV$ \cite{lum_llg}. This allows to measure the luminosity to 1\%
in a mass window $\pm 2 \GeV$ around the Higgs mass in one year of running
using muons only.  Figure \ref{fig:llg} shows the energy spectrum of
photons from the process $\gmgm \rightarrow \ell^+ \ell^- \gamma$ for
$|J_z| =2$ and $J=0$. While for $|J_z|=2$ the usual 1/E spectrum can
be seen for $J=0$ the cross section is actually rising with the photon
energy.

For $\egm$ running the situation concerning polarisation is slightly
more complicated. For weak processes the electron coupling depends on
the electron spin while all processes depend on the $\egm$
angular momentum so that in principle the luminosity for all four
possible polarisation states needs to be known. On the other hand QED
conserves parity so that with a QED process only the luminosity for a
given $\egm$ angular momentum can be measured. Fortunately the
electron and $\gamma$ polarisation at high energies are quite high and
the electron polarisation can be measured with high accuracy.

The differential cross section $\egm \rightarrow \egm$ is proportional to
$1/(1-\cos \theta)$ for $|J_z| = 1/2$ while it is proportional to 
$(1-\cos \theta)$ for $|J_z| = 3/2$. In the real $\egm$ mode, where
only one beam is converted there is thus a chance to get both
components from the angular dependence. In the parasitic mode, where
in general only $|\cos \theta|$ can be measured it seems difficult to
identify the small $|J_z| = 3/2$ component.
$\egm \rightarrow {\rm e}^- \ee$ has a measurable cross section within
the detector independent from the beam polarisation and can be used to 
measure the $|J_z| = 3/2$ component once the $|J_z| = 1/2$ component is known
\cite{lumi_telnov}.

The process $\egm \rightarrow {\rm W}^- \nu$ is sensitive to left
handed electrons only. Again the triple gauge couplings need to be
known if the process should be used for luminosity determination

\section{Technical Issues}
\label{sec:techincal}

The $\gmgm$ collider imposes several technical challenges \cite{tdr_gg}. The
most important one is certainly the laser system. In order to reach high
conversion factors a laser power of $\mathcal{O}$(10\,J/pulse) and a spot size
of $\mathcal{O}(10 \micron)$ is needed.

Due to the interaction with the laser the electron beam gets
significantly disrupted.  This requires a large crossing angle between
the beams. Due to these two facts a large energy is deposited on the
detector surface which potentially result in substantial background in
the detector.

Another challenge is the beam dump. Photons cannot be deflected, so that the
energy density at the dump is large. There will be a
direct line of sight from the dump to the interaction point so that neutrons
created in the dump can reach sensitive detectors.

Another problem is the feedback system. The disrupted beam can
probably not be used for a fast feedback, since the low energy tail is
too strongly deflected by the detector solenoid. A possible idea is to
use low charge bunches between the main bunches \cite{feedback}.
A detailed study is needed to proof that this is possible.

\newcommand{\x}{\hspace{1ex}}

\subsection{The Laser Cavity}
\label{sec:laser}

The production of 10\,J laser pulses with a frequency of 10\,kHz
is difficult if not impossible by today's standards. But from the
more than $10^{19}$ photons in a laser pulse only
$\mathcal{O}(10^{10})$ are used per beam-laser interaction. This
makes it natural to reuse a laser pulse many times. An optical
ring resonator has been proposed for this purpose
\cite{Telnov:1999tb,Telnov:1999zz,Will:2001ha}.  For this option a
conceptual design has been developed \cite{KMW05}, which is
summarised below. The time between two bunches at the ILC is 
337\,ns, so that the total length of the cavity is approximately
100\,m.

\subsubsection{Power enhancement within a passive optical cavity}

Inside an optical ring resonator the circulating electric field
directly after the input coupling mirror $M_c$ is given by
a superposition of the transmitted incoming electric field
and a scaled replica of the circulating field that
emerged from this coupling mirror at the previous round-trip.
On resonance the power enhancement
factor $A_q$ describes a monotonous increase of power after the cavity
has been filled with $q$ pulses:
\[
  A_q = (1-R_c)
 \left[
 \frac{\displaystyle 1-\left( \sqrt{R_c\,V}\right)^q}
   {\displaystyle 1-\sqrt{R_c\,V}}
 \right]^2
\]
$R_c$ represents the intensity reflectivity of the coupling mirror
and $V$ the power loss factor for one round-trip.

If the reflectivity of the coupling mirror equals the loss factor
$V$ of the cavity, the maximum possible steady-state power
enhancement factor of $A_\infty = 1/(1-V)$ is obtained and all
light is absorbed by the resonant cavity. This is known as
impedance matching.

On resonance no power will be reflected from an impedance matched
cavity in steady state. The reflected power can be used as an
indicator for the alignment of the cavity in an automated control
system.

\subsubsection{Proposed design for an optical cavity}
To maintain a sufficiently high photon density and hence a high
conversion factor, the laser pulse must be focused at the Compton
conversion point (CP). When operated at the inevitable high power
level optical windows within the cavity would imply the risk of
distortion of the circulating optical picosecond-pulse as a result
of the non-vanishing B-integral \cite{Sieg_b:1986,Koe:1992}. The
Compton-interaction requires therefore operation of the cavity in
the vacuum of the accelerator.

Due to space limitations and to avoid vibrations any component of
the optical cavity should preferably be positioned outside the
environment of the particle detector, a few metres away from the
laser focus. As a consequence of the required tight optical focus
and the large distance the cross section of the 
laser beam will increase to more than half a metre at the location of
the final focusing mirror.

\begin{figure}[htbp]
\centering
\includegraphics[width=0.9\textwidth,bb=58 24 1044 745,clip]{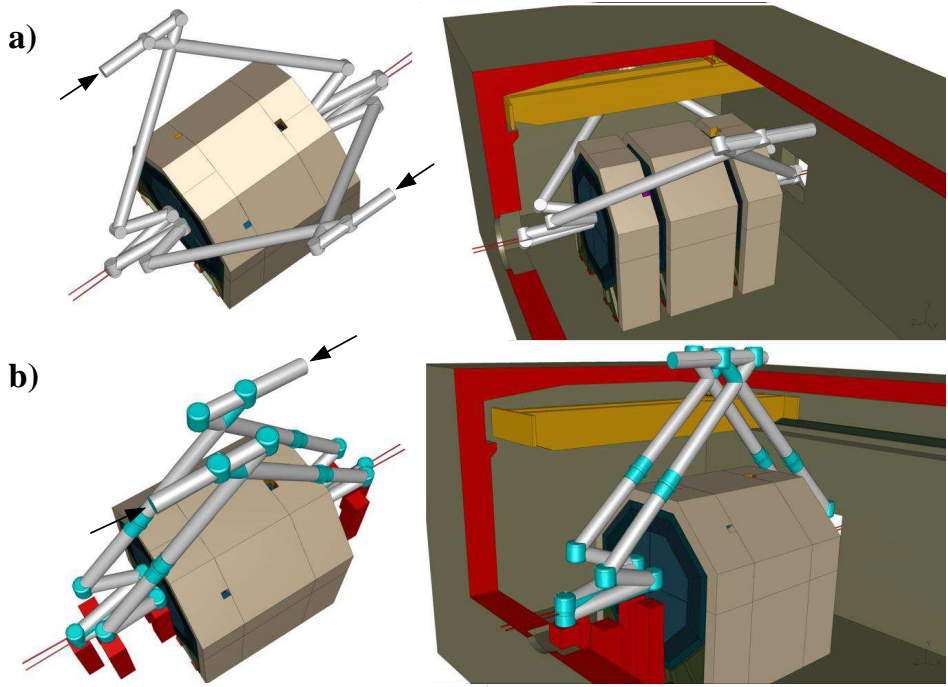}
\caption{Schematic aerial view on two possible configurations for
folding both optical cavities for the TESLA photon collider around
the detector ({\sl left}). Their respective placement in the
experimental hall is also depicted ({\sl right}). The laser beams
are coupled into the cavity at positions marked by the arrows. The
optical beam path is contained within the sketched pipes that
preserve the vacuum. The high power lasers themselves will be
located in a separate hall above the detector (not shown). The
thin lines traversing the detector represent the electron beam
paths. A slight mutual vertical tilt between the cavities permits
free passage of the particle beams. {\sl As a ruler:} The detector
extends 14.8\,m along the electron path.}
\label{fig:embedded-cavities}
\end{figure}

Compton conversion at each of the counter-propagating electron
beams requires two ring cavities as exemplified by
Fig.~\ref{fig:embedded-cavities}. They are interlaced without
mutual interference.
Their optical paths are enclosed within the associated optical
beam pipes which are needed to maintain the vacuum. For details
see \cite{KMW05}. The focusing is accomplished by means of a
mirror telescope (inverted beam expander) and a second, identical
telescope is used for re-collimation.

\begin{figure}[htbp]
  \centering
  \includegraphics[width=0.8\linewidth,bb=11 0 500 345]{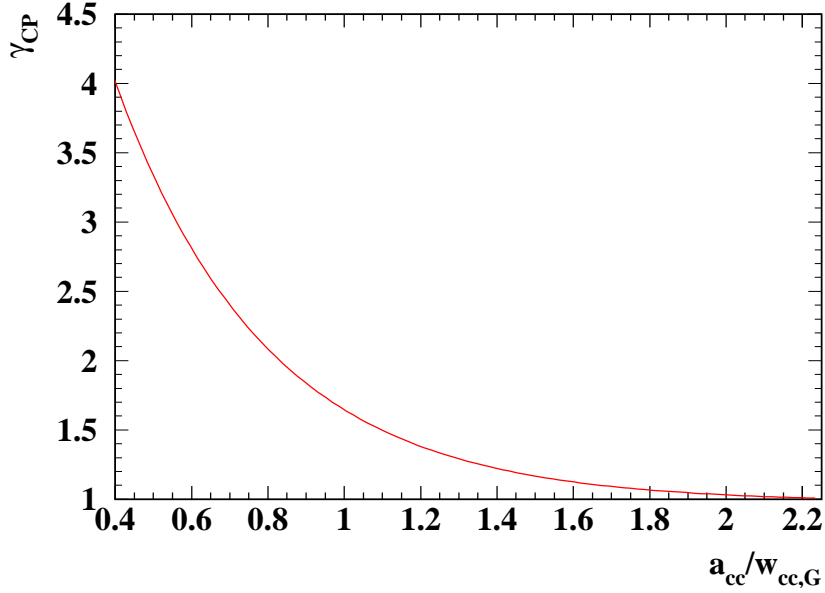}
  \caption[]{Diffraction broadening of the focal spot due to the finite size
    of the focusing mirror as a function of the radius $a_{cc}$ of
    this mirror. The x-axis shows the size of the mirror normalised
    to the Gaussian radius $w_{cc,G}$ of the beam ($1/e^2$
    convention). The y-axis shows the size of the focus normalised
    to its value for infinite mirrors.
  }
  \label{fig:dbroad}
\end{figure}

The focal spot size at the CP and the influence of the finite
diameter of the mirrors were calculated numerically \cite{KMW05}.
Total correction of the aberrations introduced by the telescope
mirrors was thereby assumed. That necessitates e.g. the use
of off-axis parabolic mirrors. The diameter of the final
focusing concave mirrors (denoted by the subscript "cc" in related
symbols) determines the minimal collision angle $\alpha_L$ between
the laser and the electron beam. It should be kept small for high
yield of photons. The major effect in reducing the size of the
mirrors is a diffractive broadening of the focal spot size. The
increased loss due to radiative energy that spills over the
boundaries of the mirrors turned out to be much smaller. Figure
\ref{fig:dbroad} shows the broadening of the focal spot due to the
finite size of the focusing mirror as a function of the radius of
this mirror.

\subsubsection{Laser-electron crossing angle}
\label{sec:lumiopt}

\begin{figure}[htb]
\centering
\includegraphics[width=\textwidth]{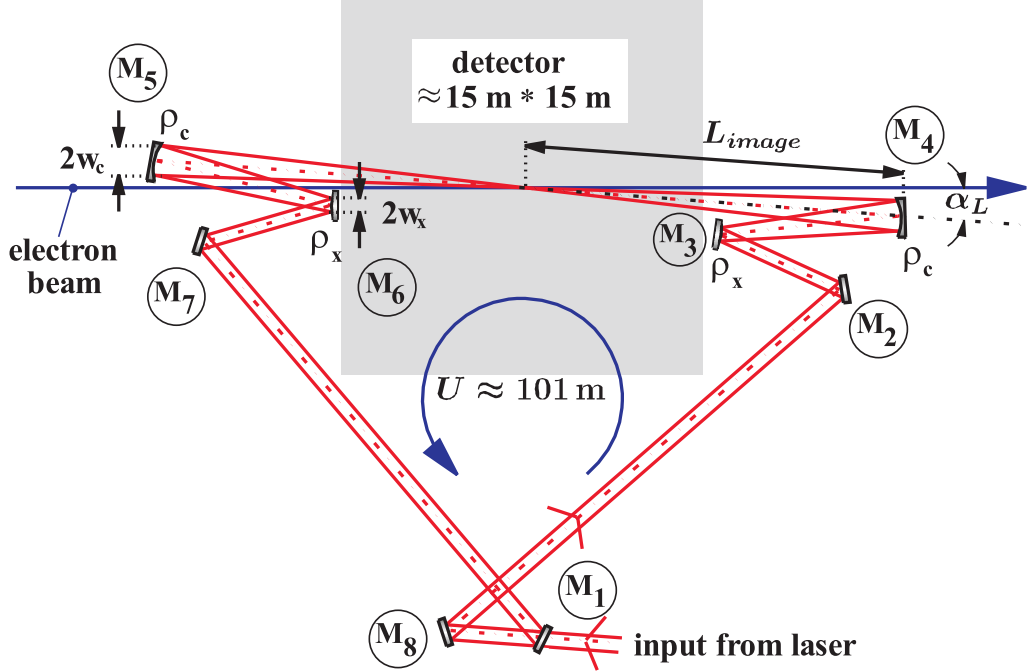}
\caption{Geometry (to scale) of one of the identical plane cavity,
comprising a beam magnification $\mu=w_c/w_x$.}
\label{fig:telescopic_cavity}
\end{figure}

In order to calculate the laser-electron crossing angle $\alpha_L$
and to specify the beam waist $\widetilde{w_0}$ for finite mirror
size, diffraction broadening has to be taken into account.  To
optimise the yield of photons, the crossing angle $\alpha_L$, the
mirror diameter $2\,a_{cc}$, the waist size $\widetilde{w_0}$, the
laser pulse energy $E_{pulse}$, as well as the laser pulse
duration $\tau_{pulse}$ are all interdependent parameters. Their
respective values were determined by a numerical optimisation
process using the program CAIN \cite{cain} to calculate the yield.
CAIN assumes that charged particles interact with a Gaussian
optical beam. The centre-of-mass energy was set to 500\,GeV.  The
aperture $2\,a_{cc}$ of the final focusing concave mirrors $M_4$,
$M_5$ at distance $L_{image}$ from the conversion point CP in
Fig.~\ref{fig:telescopic_cavity} sets an upper limit for the
opening angle $\theta_{cc}$ of the laser cone that emerges from
the beam waist:

\[
  \theta_{cc}=\frac{\displaystyle
    a_{cc}}{\displaystyle L_{image}} = \frac{\displaystyle
    a_{cc}}{\displaystyle w_{cc,G}}\, \theta \quad.
\]

\begin{figure}[htb]
\centering
\includegraphics[width=0.3\linewidth]{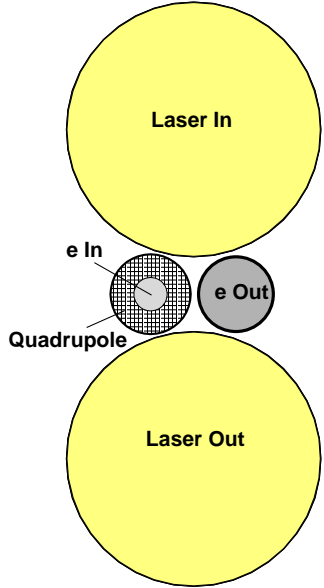}
\caption{Arrangement of the laser- and beam-pipes at the front face of
  the final quadrupole ($z = \pm 3.8\,{\rm m}$).}
\label{fig:lscetch}
\end{figure}

Replacing $L_{image}$ by the far field divergence angle $\theta$ originating
from a Gaussian beam waist $w_0$ results in the latter equation. $w_{cc,G}$
represents the beam radius on each of the concave mirrors.
If the electron beams cross each other in the horizontal ($x-z$) plane,
the laser beams should run in the vertical ($y-z$) plane. To allow for
an opening angle of the outgoing beampipe of 14\,mrad and the finite size of
the focusing quadrupole, it is assumed that the lower edge
of the laser beampipe stays away from the horizontal plane by
17\,mrad.
If the quadrupole can be smaller a lower offset angle is possible allowing for
a slightly smaller laser power or higher luminosity.
The projection of the different pipes in the $x-y$ plane at
the front face of the final quadrupole is shown in figure
\ref{fig:lscetch}.

\begin{figure}[htb] \centering
 \includegraphics[viewport=0 0 490 490,width=0.6\textwidth]{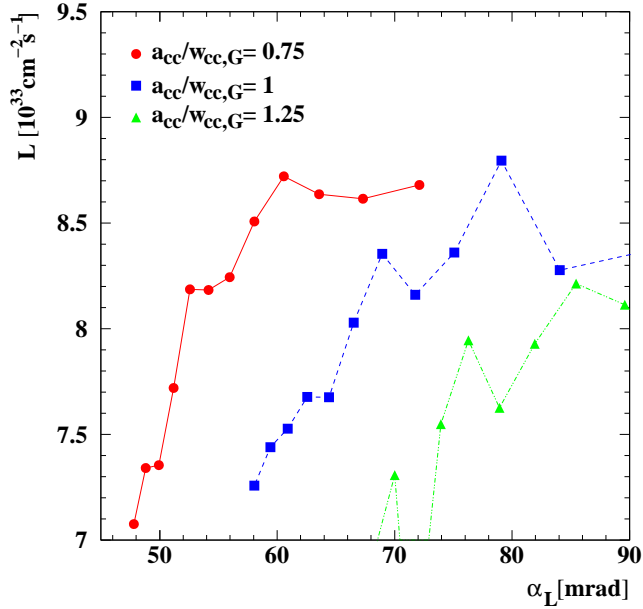}
 \caption{The $\gamma\gamma$ luminosity in
 the high energy part of the Compton spectrum as calculated using
 CAIN. It is plotted as a function of the laser-beam crossing angle
 $\alpha_L$ for different values of the normalised half-diameter
 $a_{cc}/w_{cc,G}$ of the concave mirrors. $w_{cc,G}$ denotes
 the Gaussian beam radius at this location.
 The high energy part was defined as $z>0.8\,z_{max}$, whereas
 $z_{max}=x/(x+1)$, neglecting non-linearity effects.
 \label{fig:lopt2}}
\end{figure}

The crossing angle $\alpha_L$ is thus expressed as
\[
 \alpha_L = \frac{\displaystyle
 a_{cc}}{\displaystyle w_{cc,G}}\, \theta +\beta\ , \qquad
 \beta=17\,\mbox{mrad} \quad.
\]
This relation was encoded into CAIN via the optical Rayleigh
length $z_R=\pi\,w_0^2/(M^2\lambda$) and the relation $\theta=w_0/
z_R$. The beam quality factor $M^2$ equals $1$ only for Gaussian
beams \cite{Sieg_d:1986}. For more general beams holds
$M^2$\,$>$\,1, and the waist can be expressed as
$\widetilde{w_0}$\,=\,$M^2\,w_0$. It turned out that near the
focus the diffraction broadened beam is well approximated when
substituting $M^2$ by $\gamma$, and hence
$\widetilde{w_0}$\,=\,$\gamma\,w_0$ has been used instead of the
Gaussian beam waist $w_0$. Fig.~\ref{fig:lopt2} shows the
resulting luminosity as a function of the crossing angle
$\alpha_L$ for different mirror sizes $a_{cc}/w_{cc,G}$ and a laser
pulse duration of $\tau_{L}=3.5$\,ps FWHM ($\sigma$\,=\,1.5\,ps).
In the examined range the luminosity rises with decreasing
diameter of the mirrors. A value of $a_{cc}/w_{cc,G}=0.75$ is
therefore selected. An acceptable crossing angle is then
$\alpha_L\,\approx\,55$\,mrad.  This corresponds to a Rayleigh
length $z_R$\,$\approx$\,0.63\,mm, a diffraction broadened beam
waist of $\widetilde{w_0} $\,$\approx$\,14.3\,$\mu$m\footnote{
  $\widetilde{w_0}$
   is given in the $1/e^2$ convention, designating the radius
  that is defined by a drop of the intensity to
  $1/e^2$\,$\approx$\,13.5\,\% of its maximum value at the beam
  centre. Given in Gaussian $\sigma$ its value is a factor two smaller.}
and a nominal Gaussian waist of $w_0 $\,$\approx$\,6.5\,$\mu$m.

The behaviour at small crossing angle has been verified by a program
calculating the overlap integral of the electron beam and the laser
and can be understood as follows. Because of the offset the length of
the overlap region increases with $\sqrt{z_R}$, i.e.~smaller crossing
angles. On the other hand the beam widens in two directions so that the photon
density decreases with $1/z_R$ resulting in a $1/\sqrt{z_R}$ decrease
of the conversion probability. Without the offset the length
increases in principle with $z_R$, however the length of the overlap
region is limited by the duration of the laser spot. 
At large crossing angle the overlap integral continues to increase.
However the conversion probability is limited to one and because of
the increasing non-linearity the spectrum is shifted to lower energies
which are not considered in our definition of the luminosity.

A total luminosity\footnote{Here $z_{max}$ is defined as
  $z_{max}=x/(x+1+\xi^2)$ consistent with the definition in
  \cite{tdr_gg}.} of ${\mathcal L}(z>0.8 z_{max}) = 1.1 \cdot
10^{34}\,{\rm cm^{-2}s^{-1}}$ can be achieved for these parameters
with a pulse energy of 9\,J \cite{Bech:03}. A non-linearity parameter
$\xi^2=0.3$ can be maintained in accordance with reference
\cite{tdr_gg}, however the required laser power is 80\% larger. In
proportion to the laser pulse energy the required average laser power
has also gone up to
$9\,{\rm J} \times 2820 \times 5\,{\rm Hz} \approx 130\,{\rm kW}$. All
resulting parameters for the Compton interaction zone of a
$\gamma\gamma$-collider based on 250\,GeV electron beams are compiled
in Tab.~\ref{tab:Laser-require}.

\begin{table}[htbp]
\centering
\caption{Optical parameters resulting
from an optimisation of the $\gamma\gamma$ luminosity}
\begin{tabular}{ll}\hline
laser pulse energy $E_{pulse}$        & $\approx$\,9.0\,J\\
average laser power $<P_{laser}>_t$   & $\approx$\,130\,kW\quad
for one pass
collisions at the\\
& \hspace{11.5ex}TESLA bunch-structure\\[0.5ex]
pulse duration $\tau_{pulse}$         & 3.53\,ps FWHM ($\sigma$\,=\,1.5\,ps)\\
Rayleigh length $z_R$                 & $\approx$\,0.63\,mm\\
beam waist $w_0$                      & $\approx$\,14.3\,$\mu$m\,($1/e^2$) ($\sigma$\,=\,7.15\,$\mu$m)\\
laser-e$^-$ crossing-angle $\alpha_L$ & $\approx$\,56\,mrad \\
normalised mirror-size $a_{cc}/w_{cc,G}$          & 0.75 \\
laser wavelength $\lambda$            & 1.064\,$\mu$m\\
nonlinearity parameter $\xi^2$        & 0.30 \\
total luminosity $L_{\gamma\gamma}$   & $1.1\cdot 10^{34}$\,cm$^{-2}$s$^{-1}$\\[2ex]
\hline
\end{tabular}
\label{tab:Laser-require}
\end{table}

\subsubsection{Enhancement capability of the cavity}
According to the results obtained in \cite{KMW05} all mirrors could
have a diameter of about 120\,cm. In this case cutoff occurs at
approximately 75\,\% of the hypothetical Gaussian beam radius
$(1/e^2)$ at the final focusing mirror which represents the dominant
aperture at which diffraction will occur.

An estimated fractional power loss due to diffraction of roughly
$LF_{\rm diff}\ge 0.9998$ per round trip and a reflectivity of
between $R_{HR}$\,=\,99.99\,\% and 99.95\,\% for presently
available standard mirror coatings would permit a steady-state
impedance matched power enhancement between 1100 and 270, for
otherwise perfect conditions\footnote{Such reflectivities are
currently readily available only for mirror substrates of a few cm
diameter. This is not limited due to any principles of physics.
For optical gravitational wave detection similar mirrors
of about 30\,cm diameter have already been manufactured. This
topic certainly requires further technological effort.}. The
enhancement becomes the more sensitive against any impedance
mismatch, the larger $A_\infty$ is.

The optical energy fluence of $\approx$\,13\,J/cm$^2$ is expected
to be well below the damage threshold of mirror substrates and
coatings. However, no data for trains of ps-pulses separated on
nano- to microsecond time scales which accumulate to the stated
fluence exist. For a final judgement an experimental
study with a representative of the ILC bunch structure would be
required.

\subsubsection{Effects of cavity misalignments}
Any misalignment of a mirror position and orientation generally results in
displacement and broadening of the intra-cavity beam waist.
According to our calculations, the displacement of the beam waist
remains smaller than the Rayleigh length, i.e.~the depth of the
focus. This shift of the beam waist is hence negligible.

Transversely the laser beam has to collide with the electron beam
which requires a precision of $< 10\,\micron$.  Feedback algorithms
for this need to be developed. They could use the power of the high
energy photon beam as well as the optical radiation leaking through
one of the focusing mirrors around the Compton interaction point

Maintaining the power enhancement factor e.g. above 90\,\% of its
optimum value $A_\infty$ demands sub-nm precision for controlling
the circumference. Technical solutions for such a precise length
stabilisation are well-known \cite{DH:1983}. Even more stringent
restrictions are common in interferometrical detection of
gravitational waves. For the latter case control loops for
automatic alignment have been developed. Despite the different
operation modes, the use of bursts of optical ps-pulses for the
$\gamma\gamma$ collider and continuous-wave (cw) laser radiation
in optical interferometers for gravitational waves, the cavity for
the $\gamma\gamma$-collider should benefit from that knowledge. In
addition, adaptive optics appears to be essential for operation of
such a cavity for the photon collider \cite{KMW05}.


\subsection{Detector and Backgrounds}
\label{sec:detect}
\begin{figure}[htbp]
  \centering
  \includegraphics[width=0.6\linewidth,bb=8 3 502 467]{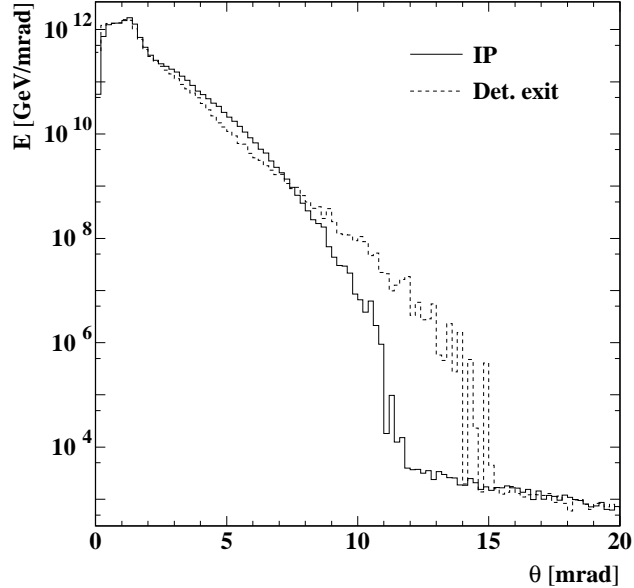}
  \caption{Energy weighted angular distribution of the outgoing electron beam
    directly behind the interaction point (solid) and at $z=2.8\,{\rm m}$
    (dashed).}
  \label{fig:diff}
\end{figure}

Machine backgrounds at a linear collider are mainly coming from beam-beam
effects in the interaction region \cite{tdr_det}. There are
several differences between an $\ee$ and a $\gmgm$ collider: The
electron beam gets disrupted in the electron-laser interaction with a
disruption angle around 10 mrad. 
This precludes head on collision
since the outgoing beam does no longer fit through the aperture of the
final focusing quadrupole. 
The beam radius increases further due to the angle between the 
outgoing beam and the magnetic field of the detector solenoid. Figure
\ref{fig:diff} shows the angular spread of the outgoing electron beam directly
behind the interaction point and and at $z = 2.8 \, {\rm m}$.
To limit the energy loss in the detector an opening angle of 14\,mrad for
the outgoing beampipe has been chosen.
A crossing angle of 34\,mrad
had been adapted, as suggested in \cite{tdr_gg}. The front face of the
final quadrupole is assumed to be $l^* = 3.8 \, {\rm m}$ away from the
interaction point. With recent designs of small superconducting quadrupoles
\cite{parker} this crossing angle or even a slightly smaller one should be
possible. 

\begin{figure}[htb]
  \centering
  \includegraphics[width=0.7\linewidth]{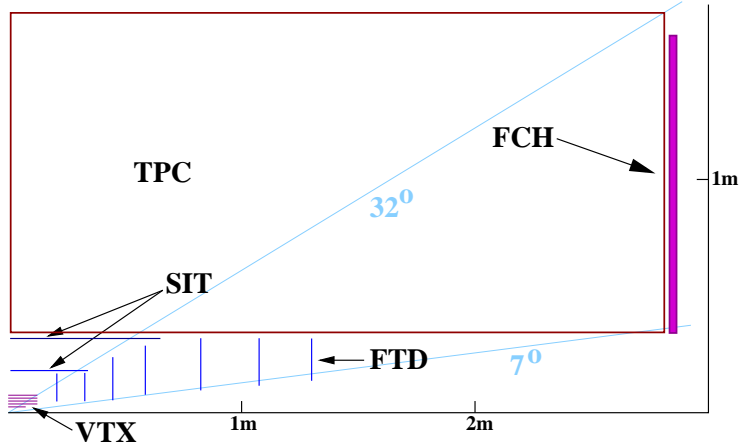}
  \caption{Tracking system of the TESLA TDR detector.}
  \label{fig:track}
\end{figure}

For simplicity, the detector is assumed to be identical to the $\ee$
detector described in \cite{tdr_det} above polar angles of $\theta =
7^\circ$. Below this angle it has to be modified to accommodate the
$\gmgm$ specific equipment. The tracking system of the detector is
shown in figure \ref{fig:track}.

Detailed backgrounds have been simulated using CAIN \cite{cain}. For
these simulations the incoherent particle-particle as well as the
coherent particle-beam interactions have been considered. The direct
background at large angle is exclusively coming from incoherent $\ee$
pair creation. This background is significantly smaller than in the
$\ee$ version of TESLA \cite{tdr_det}. Another significant amount of
background is coming from backscattering of particles. This
background is potentially larger than in $\ee$ for several
reasons. Because of the crossing angle the particles already hit the
detector at a larger radius than in $\ee$. Also because of the
crossing angles the beams are not parallel to the solenoid, so that
low momentum particles get swept out of the beampipe by the magnetic
field. Low energy electrons of one bunch get deflected by the negative
charge of the opposing bunch, while they are focused by the opposite
charge in $\ee$. Some electrons get deflected enough by the
electron-laser interaction that they hit the detector or the inner side
of the beampipe close to the detector.

\begin{figure}[htb]
\centering
\includegraphics[width=0.7\linewidth,bb=10 3 565 440]{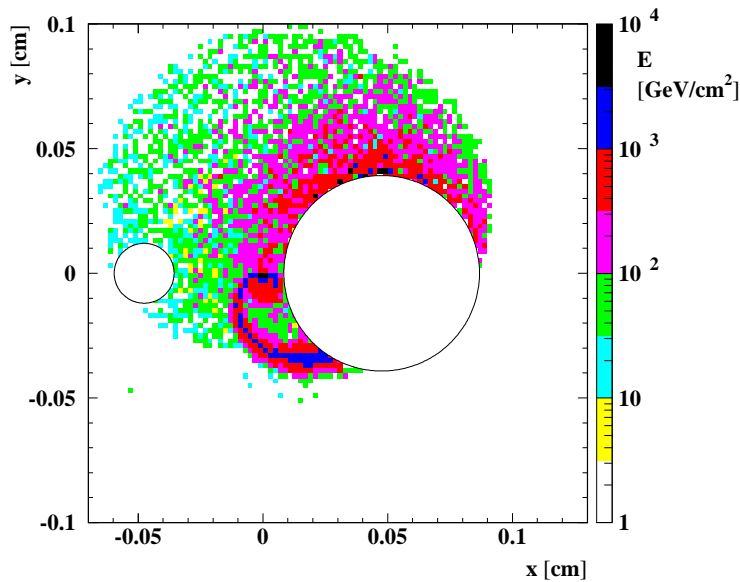}
\caption{Energy deposition from one bunch crossing at $z=2.8\,{\rm m}$, the
  front face of the ECAL.}
\label{fig:ggbck}
\end{figure}

Figure \ref{fig:ggbck} shows the energy distribution at $z=2.8\,{\rm m}$, 
the front-face of the
electromagnetic calorimeter.
In total 40\,TeV per bunch crossing are deposited on the front face of
the mask from pair production at the IP. This is roughly the same
energy as for the $\ee$ option described in \cite{tdr_det}. Because of
the large crossing angle, however, the backscattered particles are more
difficult to capture by the masking system. There is also a danger
that several hundred TeV of electrons hit the mask very close to the
exit whole. These electrons stem from multiple interactions with the
laser beam. They have a significantly larger energy than the pairs and
contribute thus much less to the backscattering. If needed these
particles can be suppressed by making the beampipe in the horizontal
plane slightly larger.

The inner part of the detector had to be completely redesigned to house the
pipes for the beam and the laser. To avoid scattering surfaces inside the
detector the beampipe is identical to the $\ee$ case up to $z=17\,\cm$
followed by a conical part with opening angle 93\,mrad up to $z= \pm 2.825
\,{\rm m}$.

Only behind the front-face of the calorimeter the common beampipe
splits into individual ones for the incoming and outgoing beams and
lasers.  To absorb backscattered particles the pipes are surrounded by
a tungsten mask with pointing geometry and at thickness of 5\,cm at
$z=2.8\,{\rm m}$.

\begin{figure}[htbp]
  \centering
  \includegraphics[width=0.7\linewidth]{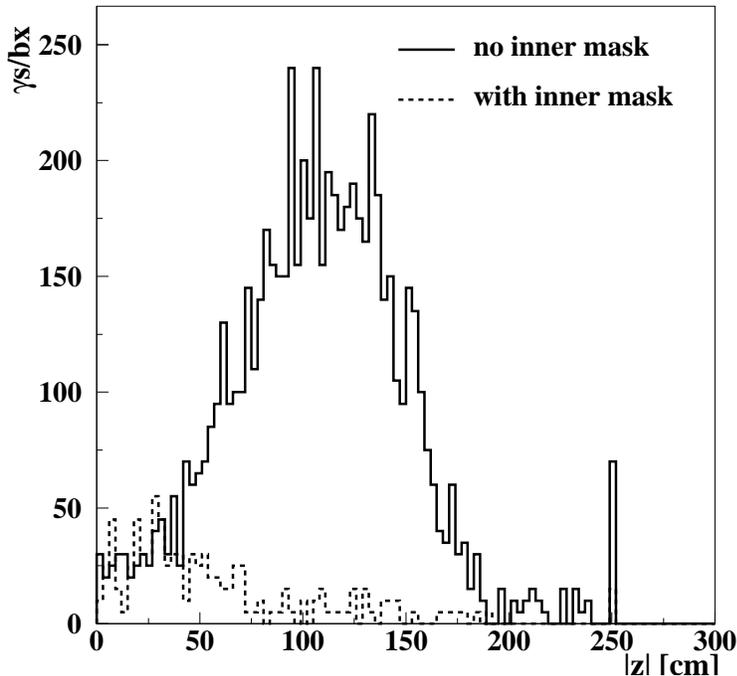}
  \caption{$z$-coordinate of photons entering the TPC.}
  \label{fig:masktpc}
\end{figure}

Another mask of 5\,cm thickness is put inside
the beampipe where the space needed for the laser is left free.
Since the background is not symmetric in the azimuthal angle the function of
the mask is not deteriorated by the missing pieces.
The outer mask starts at $z$=23\,cm to protect as much as possible of
the detector,
the inner mask starts at $z$=1\,m because otherwise it receives too many direct
hits from background particles created at the interaction point.
The photon background in the TPC as a function of the $z$-coordinate where the
photon enters the TPC is shown in fig.~\ref{fig:masktpc} with and without the
inner mask. It is evident that the inner mask is needed to protect the TPC.

\begin{figure}[htb]
\centering
\includegraphics[width=0.7\linewidth]{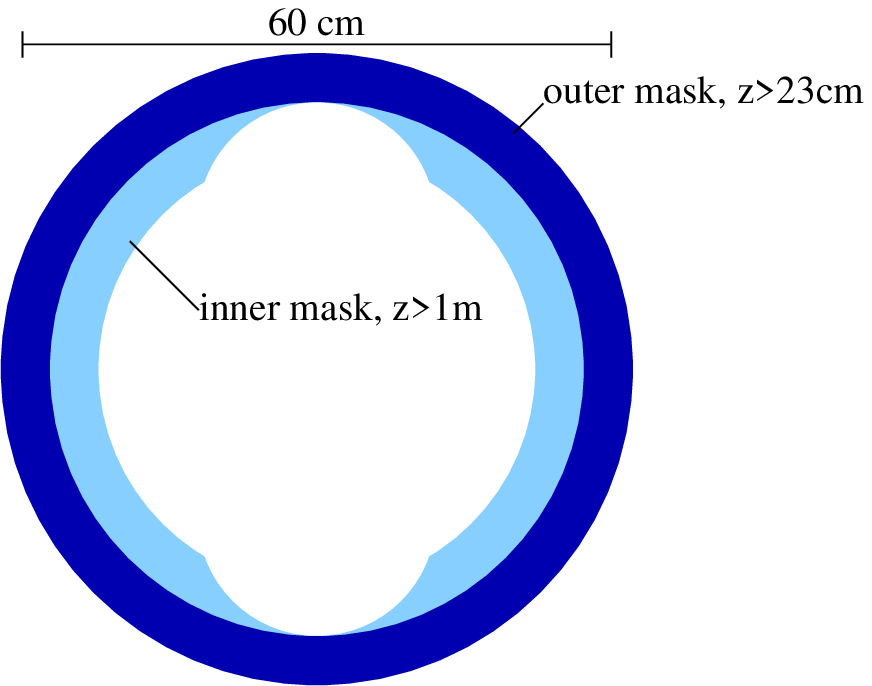}
\includegraphics[width=\linewidth]{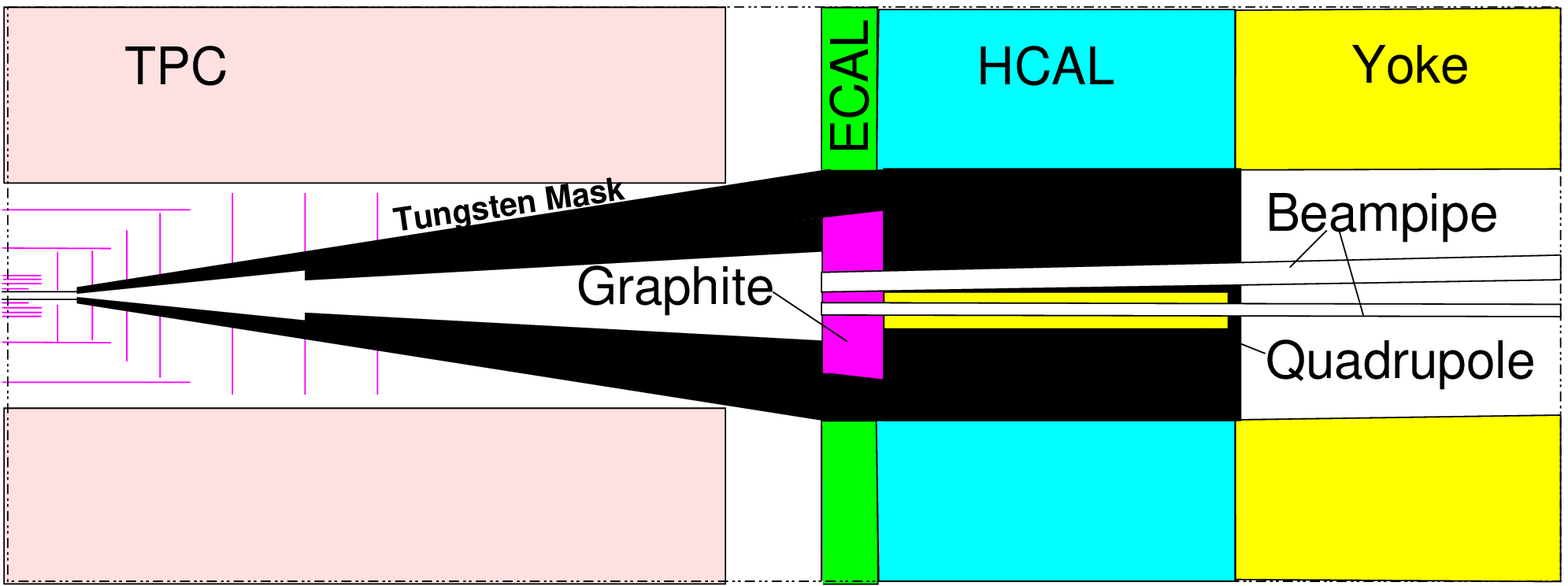}
\caption{$x-y$ projection at $z=2.8\,{\rm m}$ (upper plot) and $x-z$ 
  projection of the inner region of the $\gmgm$-detector.}
\label{fig:mask}
\end{figure}

Figure \ref{fig:mask} shows the beampipe region in the $x-y$ and $x-z$
projection.
This setup has been simulated using the TESLA simulation program BRAHMS
\cite{brahms} which is based on GEANT3 \cite{geant3}.
The detector is hermetic above $\theta = 7^\circ$. Inside the mask it
should be possible to install some tagging devices for photon
structure function measurements left and right of the beampipe (see
fig. \ref{fig:lscetch}) where the background level is relatively low
(fig. \ref{fig:diff}). In $\egm$ running it should also be possible to
replace the unused laser pipe at negative $y$ by a tagging device.

\begin{figure}[htb]
  \centering
  \includegraphics[width=0.5\linewidth,bb=11 2 493 474]{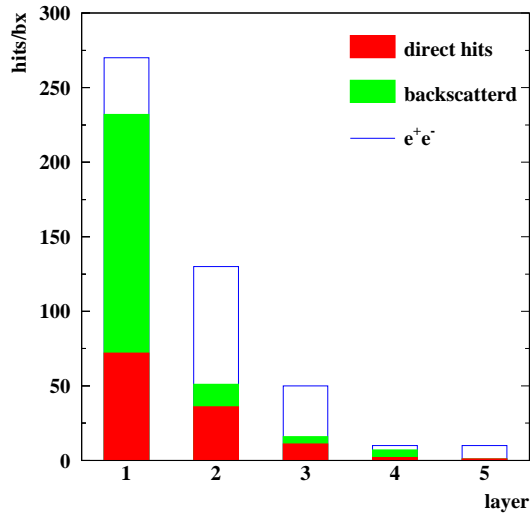}
  \caption{Hits in the microvertex detector from $\ee$ pairs created in
    the interaction region.}
  \label{fig:bgvd}
\end{figure}

Figure \ref{fig:bgvd} shows the background in the vertex detector separated
into direct hits and hits from backscattering. The total number of hits in the
different layers is very similar to $\ee$, however in $\ee$ basically all hits
are direct. In $\gmgm$ the innermost layer of the vertex detector cannot be
protected by the mask so that it receives a significant amount of
backscattering. 
The background shown in figure \ref{fig:bgvd} is only from incoherent
processes at the interaction point. 
In addition a background of two times around 15 hits per bunch
crossing and layer is present from low energy $\gmgm \rightarrow \qq$
events, explained in section \ref{sec:pileup}, and from backscattering
from showers induced by electrons that loose exceptionally much energy
in the interaction with the laser or by beamstrahlung. The latter
process is very rare, but consist of relatively large showers, so that
the fluctuations are large.

The number of photons passing the TPC is estimated to be 1800/bunch crossing.
Also this number is comparable to the $\ee$ case
\cite{tdr_det} and should be manageable.

\subsection{The Beam-dump}
\label{sec:dump}
Since photons cannot be deflected there has to be a direct line of sight from
the interaction point to the beam dump. The standard ILC design uses a
water dump at a distance of about 100\,m from the detector. Such a dump has
been simulated with Geant4 \cite{geant} using the physics list QGSP\_HP and a
cross section bias of 100.  Cross checks have been done with LHEP\_GN,
QGSP\_GN, and a cross section bias of one and consistent results have been
found.  For a neutron kinetic energy of $E_n > 15 \keV$ about 3.5
neutrons/BX/cm$^2$ have been found with this setup from the $\gamma$-beam only,
corresponding to $5\cdot 10^{11}$ neutrons/cm$^2$/year. If the electron beam
will be dumped in the same beam dump this number has to be doubled.  Such a
neutron flux will be a problem for a CCD vertex detector. Some ideas how to
reduce the neutron flux exist \cite{valery_dump}, but there is no detailed
design yet.

Another problem for the dump is the high energy density of the photon beam
which would heat the water in the dump locally too much. Some ideas how to
avoid this are also presented in \cite{valery_dump}.


\section{Conclusions}
\label{sec:conc}

Depending on the physics scenario, nature has chosen, a photon collider is an
interesting complement to the $\ee$ baseline version of the ILC. The
final decision if a photon collider will be built, should only be taken when
supporting results from LHC and $\ee$ running of the ILC are available.

From the technical side the greatest challenge is probably the laser system. A
conceptual design for a resonant laser cavity has been shown which could reach
a power enhancement factor around 100 and thus reduce the required laser power
to an acceptable level.

The detector at a photon collider seems manageable. The region below a polar
angle of $7^\circ$ is completely taken by pipes and the masking system, but at
larger angles a detector similar to $\ee$ can provide comparable performance.

\section*{Acknowledgements}
This work profited from useful discussion with many people. In particular we
wish to thank
G.~Franzoni,
J.~Gronberg,
W.~Kilian,
F.~Krauss,
A.~Leuschner,
J.~List,
N.~Meyners,
D.~Miller,
V.~Telnov,
N.~Walker,
I.~Will,
K.~Zapfe and
A.~Zarnecki.

\end{document}